\newif\ifIEEE
\newif\ifPAGELIMIT
\newif\ifSIAM
\newif\ifJCTA
\PAGELIMITfalse
\IEEEfalse
\JCTAtrue
\SIAMtrue  
\ifPAGELIMIT
    \IEEEtrue
\fi
\ifIEEE
    \JCTAfalse
    \SIAMfalse
\fi
\ifJCTA
    \SIAMfalse
\fi
\ifIEEE
    \ifPAGELIMIT
        \documentclass[conference,final,letterpaper]{IEEEtran}
    \else
        \documentclass[journal,final,letterpaper]{IEEEtran}
    \fi
    \newcommand{\bibauthor}[1]{#1}
    \newcommand{\bibpaper}[1]{``#1''}
    \newcommand{\Footnotetext}[2]
    {
        \begin{figure}[!b]
        \footnotesize\vspace{-3ex}\hrulefill\hfill
        \makebox[0em]{}\hfill\makebox[0em]{}\par${}^{#1}$ %
                                                        #2\vspace{-.6ex}
        \end{figure}
        \addtocounter{figure}{0}
    }
    \usepackage{amsthm}
\else
    \ifJCTA
        \documentclass{elsarticle}
        \newenvironment{IEEEkeywords}{\begin{keyword}}{\end{keyword}}
        \usepackage{amsthm}
    \else
        \ifSIAM
            \documentclass{siamart190516}
            \newenvironment{IEEEkeywords}{\begin{keywords}}%
                                                        {\end{keywords}}
        \else
            \documentclass[12pt]{article}
            \newenvironment{IEEEkeywords}{\begin{small}%
                                  \textbf{Index Terms} ---}{\end{small}}
            \usepackage{amsthm}
        \fi
   \fi
   \newcommand{\bibauthor}[1]{\textsc{#1}}
   \newcommand{\bibpaper}[1]{\textsl{#1}}
\fi
\usepackage{amsfonts,bm}
\usepackage{amssymb}
\usepackage{amsmath}
\usepackage{pict2e}
\newcommand{\bibbook}[1]{\textit{#1}}

\newcommand{\bibperiodical}[1]{\textit{#1}}

\ifPAGELIMIT
\fi
\ifSIAM
    \newtheorem{problem}{Problem}
    \newtheorem{remark}{Remark}
\else
    \newtheorem{theorem}{Theorem}
    \newtheorem{proposition}[theorem]{Proposition}
    \newtheorem{lemma}[theorem]{Lemma}

    \newtheorem{problem}{Problem}
    \theoremstyle{remark}
    \newtheorem{remark}{Remark}
\fi
\renewcommand{\mathbf}[1]{{\bm{#1}}}     

\newcommand{\bigcupdot}%
                 {{\textstyle{\bigcup\!\!\!\!\!\hspace{0.25ex}\cdot\;}}}
\newcommand{\calA}{{\mathcal{A}}}
\newcommand{\calB}{{\mathcal{B}}}
\newcommand{\calD}{{\mathcal{D}}}
\newcommand{\calE}{{\mathcal{E}}}
\newcommand{\I}{{\mathcal{I}}}
\newcommand{\calJ}{{\mathcal{J}}}
\newcommand{\calH}{{\mathcal{H}}}
\newcommand{\calK}{{\mathcal{K}}}
\newcommand{\calL}{{\mathcal{L}}}
\newcommand{\calM}{{\mathcal{M}}}
\renewcommand{\O}{{\mathcal{O}}}
\newcommand{\calP}{{\mathcal{P}}}
\newcommand{\calQ}{{\mathcal{Q}}}
\newcommand{\calS}{{\mathcal{S}}}
\newcommand{\calT}{{\mathcal{T}}}
\newcommand{\calX}{{\mathcal{X}}}
\newcommand{\Set}{{\mathcal{U}}}
\newcommand{\Setalt}{{\mathcal{V}}}
\newcommand{\blda}{{\mathbf{a}}}
\newcommand{\bldh}{{\mathbf{h}}}
\newcommand{\bldk}{{\mathbf{k}}}
\newcommand{\bldtheta}{{\mathbf{\vartheta}}}
\newcommand{\ndiv}{\tau}
\newcommand{\hist}{{\mathbf{r}}}
\newcommand{\Prob}{{\mathsf{Prob}}}
\newcommand{\Expected}{{\mathbb{E}}}
\newcommand{\Variance}{{\mathrm{Var}}}
\newcommand{\X}{{\mathsf{\Omega}}}
\newcommand{\F}{{\mathbb{F}}}
\newcommand{\GF}{{\mathrm{GF}}}
\newcommand{\degb}{h}
\newcommand{\N}{N}
\newcommand{\Type}[1]{{\mathrm{T}(#1)}}
\newcommand{\Integers}{{\mathbb{Z}}}
\newcommand{\floor}[1]{\left\lfloor{#1}\right\rfloor}
\newcommand{\ceil}[1]{\left\lceil{#1}\right\rceil}
\newcommand{\tib}{{\tilde{b}}}
\newcommand{\tir}{{\tilde{r}}}
\newcommand{\tis}{{\tilde{s}}}
\newcommand{\tiw}{{\tilde{w}}}
\newcommand{\mult}{{\mathsf{mult}}}
\newcommand{\Comp}{{\overline{\calP}}} 
\newcommand{\random}{{\mathsf{T}}}
\ifPAGELIMIT
    \newcommand{\Withoutlossofgenerality}{W.l.o.g.}
    
    \newcommand{\respectively}{resp.}
    \newcommand{\Theorem}{Thm.}
    \newcommand{\Theorems}{Thms.}
    \newcommand{\Proposition}{Prop.}
    \newcommand{\Propositions}{Props.}
    \newcommand{\ifandonlyif}{iff}
\else
    \newcommand{\Withoutlossofgenerality}{Without loss of generality}
    
    \newcommand{\respectively}{respectively}
    \newcommand{\Theorem}{Theorem}
    \newcommand{\Theorems}{Theorems}
    \newcommand{\Proposition}{Proposition}
    \newcommand{\Propositions}{Propositions}
    \newcommand{\ifandonlyif}{if and only if}
\fi

\newcommand{\Title}{On the Number of Factorizations of
                                         Polynomials over Finite Fields}
\newcommand{\Namea}{Rachel N. Berman}
\newcommand{\Nameb}{Ron M. Roth}
\newcommand{\Address}{Computer Science Department,
                      Technion,
                      Haifa 3200003, Israel}
\newcommand{\Emaila}{rachelinka@gmail.com}
\newcommand{\Emailb}{ronny@cs.technion.ac.il}
\newcommand{\Grant}{This work was supported in part by Grants 1396/16
                    and 1713/20 from the Israel Science Foundation.
                    \ifPAGELIMIT
                    \else
                    A preliminary abstract version of this work
                    was presentated at
                    the IEEE Int'l Symposium on Information
                    Theory (ISIT), June 2020.
                    \fi
}
\newcommand{\Thnxab}{\Namea\ and \Nameb\ are with the \Address.
                    \par
                    Emails:
                    \Emaila, \Emailb}
\begin{document}
\ifIEEE
   \title{\Title}
       \author{\IEEEauthorblockN{\Namea\ \quad\quad \Nameb}\\
               \IEEEauthorblockA{\Address.\\ \Emaila, \Emailb}
       }
\else
   \ifJCTA
       \title{\Title\tnoteref{t1}}
       \tnotetext[t1]{\Grant}
       \author{\Namea}
       \ead{\Emaila}
       \author{\Nameb}
       \ead{\Emailb}
       \address{\Address}
   \else
       \title{\textbf{\Title}\thanks{\Grant}}
       \author{\textsc{\Namea}\thanks{\Thnxab} \and
               \textsc{\Nameb}\footnotemark[2]
       }
   \fi
\fi
\ifJCTA
\else
    \maketitle
\fi


\begin{abstract}
Motivated by coding applications,
two enumeration problems are considered:
the number of distinct divisors of a degree-$m$ polynomial
over $\F = \GF(q)$, and
the number of ways a polynomial can be written as a product
of two polynomials of degree at most~$n$ over~$\F$.
For the two problems, bounds are obtained on the maximum
number of factorizations,
and a characterization is presented
for polynomials attaining that maximum.
Finally, expressions are presented for the average
and the variance of the number of factorizations,
for any given $m$ (\respectively, $n$).
\end{abstract}

\ifPAGELIMIT
\else

\ifSIAM
    \fi
\fi

\ifIEEE
   \Footnotetext{\quad}{\Grant}
\fi
\ifJCTA
    \maketitle
\fi

\section{Introduction}
\label{sec:introduction}

Throughout this work, we fix~$\F$ to be a finite field of size~$q$.
Let $\F[x]$ be the set of polynomials over $\F$
and $\calM_n = \calM_n(q)$ (\respectively, $\calP_n = \calP_n(q)$)
be the set of all monic polynomials of degree
exactly (\respectively, at most) $n$ in $\F[x]$.

Given $m \in \Integers^+$ and $s(x) \in \calP_m$,
let $\ndiv(s)$ be
the number of distinct divisors of $s(x)$ in $\calP_m$ and define
\begin{equation}
\label{eq:unconstrained}
\Upsilon_m = \Upsilon_m(q) = \max_{s(x) \in \calP_m} \ndiv(s) .
\end{equation}
It is easy to see that the maximum is attained only when $\deg s = m$.
Accordingly, we say that $s(x) \in \calM_m$
is \emph{maximal} if $\ndiv(s) = \Upsilon_m$.
Given $(n,n') \in \Integers^+ \times \Integers^+$ and
$s(x) \in \calP_{n+n'}$, an \emph{$(n,n')$-factorization} of $s(x)$ is
an ordered pair $(u(x),v(x)) \in \calP_n \times \calP_{n'}$ such that
$s(x) = u(x)\cdot v(x)$.
The number of distinct $(n,n')$-factorizations of $s(x)$
will be denoted by $\ndiv_{n,n'}(s)$ and we define
\ifPAGELIMIT
    \[
    \Upsilon_{n,n'} = \Upsilon_{n,n'}(q)
    = \max_{s(x) \in \calP_{n+n'}} \ndiv_{n,n'}(s) .
    \]
    We will assume hereafter that
\else
\begin{equation}
\label{eq:constrained}
\Upsilon_{n,n'} = \Upsilon_{n,n'}(q)
= \max_{s(x) \in \calP_{n+n'}} \ndiv_{n,n'}(s) .
\end{equation}
We will limit ourselves in this work to the case
\fi
$n = n'$ and abbreviate the notation
$\ndiv_{n,n}(s)$ by $\ndiv_n(s)$.
We say that $s(x) \in \calP_{2n}$ is \emph{$n$-maximal} if
$\ndiv_n(s) = \Upsilon_{n,n}$.
Clearly, for all $s(x) \in \calP_{2n}$ we have
$\ndiv_n(s) \le \ndiv(s)$, therefore
$\Upsilon_{n,n} \le \Upsilon_{2n}$.

In this
\ifPAGELIMIT
    work
\else
paper,
\fi
we address two related combinatorial problems.

\begin{problem}[Ordinary factorization]
\label{prob:unconstrained}
Given $m \in \Integers^+$,
compute $\Upsilon_m$ and
characterize the maximal polynomials in $\calM_m$.
\end{problem}

\begin{problem}[$(n,n)$-factorization]
\label{prob:constrained}
Given $n \in \Integers^+$,
compute $\Upsilon_{n,n}$ and
characterize the $n$-maximal polynomials in $\calP_{2n}$.
\end{problem}

In particular, we show in Section~\ref{sec:lower-upper-bounds} that
\begin{equation}
\label{eq:maxNumDivisors}
\Upsilon_m = 2^{(m/\log_q m)(1 \pm o_m(1))} ,
\end{equation}
where $o_m(1)$ stands for an expression that goes to~$0$ as
$m\rightarrow\infty$,
and that essentially the same expression holds for
$\Upsilon_{n,n}$:
\begin{equation}
\label{eq:maxNumDivisors(n,n)}
\Upsilon_{n,n} = 2^{(2n/\log_q n)(1 \pm o_n(1))} .
\end{equation}
A characterization of an ($n$-)maximal polynomial will be
given in Sections~\ref{sec:characterization-unconstrained}
and~\ref{sec:characterization-constrained}.

For both problems, we also present
in Section~\ref{sec:average-case}
average case counterparts, and, \emph{inter alia},
we compute the expectations and bound the variances of
$\ndiv(s)$ and $\ndiv_n(s)$,
when $s(x)$ is drawn with respect to a particular uniform distribution
defined precisely for each of
the two problems in Section~\ref{sec:results}.

The counterpart of Problem~\ref{prob:unconstrained}
for \emph{integers} is classical and was studied over 100 years
ago~\cite[\S4]{AE},\cite{Nicolas},\cite{Ramanujan}.
Polynomial factorization over finite fields, on the other hand,
has hardly been
\ifPAGELIMIT
    considered.
\else
considered, to the best of our knowledge.
\fi
The enumeration of ordinary factorizations
was investigated by Piret in~\cite{Piret84} for $q = 2$.
Specifically, he proved that
$\Upsilon_m(2) \le (81/16)^{(m/\log_2 m)(1 + o_m(1))}$,
as part of an analysis that shows that most
binary shortened cyclic codes approach the Gilbert--Varshamov
\ifPAGELIMIT
    bound.
\else
bound (an earlier result by Kasami~\cite{Kasami} showed this only
for codes whose generator polynomials are irreducible over $\GF(2)$).
\fi
Enumeration of $(n,n)$-factorizations (Problem~\ref{prob:constrained})
is related to another coding problem, namely, the list decoding of
a certain type of rank-metric codes~\cite[\S4]{Roth18}.
In recent years, there has been a growing interest
in rank-metric codes~\cite{KK},\cite{SKK}
and, in particular, in their list-decoding
performance~\cite{Ding},\cite{RavivWachterZeh},\cite{Roth18}.
The value $\Upsilon_{n,n}$
and the expected number of $(n,n)$-factorizations of a random polynomial
in $\calP_{2n}$ are, respectively,
the largest and average list sizes of a list decoder for
the rank-metric code of $(n+1) \times (n+1)$ arrays that was considered
in~\cite{Roth18},
when the minimum rank distance is~$2$ and the decoding radius
is~$1$. It was shown in~\cite{Roth18} that for large fields
(namely, $q \ge 2n-1$), the list size is $4^{n - o_n(1)}$,
but no analysis was carried out when the field size is
\ifPAGELIMIT
    small.
\else
small
(e.g., $q$ is fixed as $n$ grows).
\fi
In addition to these coding applications, we believe that our study
of the structure of ($n$-)maximal polynomials is of independent
mathematical interest.
Our results demonstrate both similarities and differences between
Problems~\ref{prob:unconstrained} and~\ref{prob:constrained}.

\ifPAGELIMIT
\else
Turning to the average-case analysis,
Knopfmacher \emph{et al.} computed in~\cite{KKW}
the average and variance of the length of
all ordered and unordered factorizations
of polynomials in $\calM_m$ (where the length
is the number of factors occurring in the factorization).
Their analysis makes use of
the bivariate generating function of the number, $\bar{F}(m,k)$,
of ordered factorizations of polynomials in $\calM_m$
into exactly~$k$ factors~\cite[p.~196]{KKW}.
Thus, the expression for the expectation of $\ndiv(s)$ over
all $s(x) \in \calM_n$ can be easily obtained from their analysis;
nevertheless, we will include a (very short) proof for completeness.

In the next section, we summarize the results of our work.
\fi
Hereafter, $[\ell:k]$ denotes the set
$\left\{ i \in \Integers \,:\, \ell \le i \le k \right\}$.

\section{Summary of results}
\label{sec:results}

\emph{Bounds on $\Upsilon_m$ and $\Upsilon_{n,n}$.}
Our first set of
\ifPAGELIMIT
    results
\else
results,
which we prove in Section~\ref{sec:lower-upper-bounds},
\fi
includes bounds on the values of
$\Upsilon_m$ and $\Upsilon_{n,n}$.
\ifPAGELIMIT
\else
To this end, we will prove first some basic structural properties
of maximal polynomials.
We introduce next some notation
that will be used throughout this paper.
\fi

Fix an ordering $(p_i(x))_{i=1}^\infty$ on the monic irreducible
polynomials over~$\F$ which is non-decreasing in degree and denote
\ifPAGELIMIT
    $d_i = \deg p_i$.
\else
$d_i = \deg p_i$
(so we have $d_i \le d_{i+1}$ for each $i \in \Integers^+$).
\fi
Given a monic $s(x) \in \F[x]$,
let $s(x) = \prod_{i=1}^t p_i(x)^{r_i}$
be its irreducible factorization over~$\F$,
where $r_i = \mult_{p_i}(s)$
\ifPAGELIMIT
\else
is the multiplicity of $p_i(x)$
\fi
and $r_t > 0$ (thus $r_i = 0$ for every $i > t$).
We will write $\hist(s) = (r_1 \; r_2 \; \ldots \; r_t)$
and define
\[
\rho(s) = \max_{i \in \Integers^+ \,:\, d_i = 1} r_i
\;\; = \;\; \max_{i=1}^q r_i .
\]
It is easy to see that
\begin{equation}
\label{eq:factorization}
\ifPAGELIMIT
    \textstyle
\fi
\ndiv(s) = \prod_{i=1}^t (r_i + 1) .
\end{equation}

The next three propositions present basic structural properties of
maximal
\ifPAGELIMIT
    polynomials.
\else
polynomials that we prove in
Section~\ref{sec:lower-upper-bounds}.
\fi

\begin{proposition}
\label{prop:nohole}
Let $s(x) \in \calM_m$ be maximal
and let $\hist(s) = (r_i)_{i=1}^t$.
For any $i, j \in [1:t]$,
if $d_i > d_j$ then $r_i \le r_j$.
\end{proposition}

\ifPAGELIMIT
    Thus, we may assume hereafter
\else
As a consequence of \Proposition~\ref{prop:nohole}, from here onwards
we may assume (possibly with a different ordering of
the monic irreducible polynomials which is non-decreasing in degree)
\fi
that if $s(x)$ is maximal,
then $\hist(s) = (r_i)_{i=1}^t$ is all-positive.

\ifPAGELIMIT
\else
The next proposition
relates the degree $d_i$ to the multiplicity $r_i$
of any irreducible factor of a maximal polynomial $s(x)$,
in terms of the value of~$\rho(s)$
(the latter value, in turn, will be determined
in \Proposition~\ref{prop:unconstrained-rmaxexact} below).
\fi

\begin{proposition}
\label{prop:unconstrained-lowdegrees}
Let $s(x) \in \calM_m$ be maximal
and let $\hist(s) = (r_i)_{i=1}^t$ and $\rho = \rho(s)$.
For every $i \in [1:t]$:
\begin{equation}
\label{eq:rmax}
\frac{\rho + 1}{r_i + 2} \le d_i < \frac{\rho + 1}{r_i} .
\end{equation}
\ifPAGELIMIT
    Equivalently:
    $r_i \in \left\{\floor{\rho/d_i},\ \floor{\rho/d_i}-1 \right\}$.
\else
Equivalently,
\begin{equation}
\label{eq:rmax-alt}
r_i \in \left\{\floor{\frac{\rho}{d_i}},\
\floor{\frac{\rho}{d_i}}-1 \right\} .
\end{equation}
Moreover, (\ref{eq:rmax-alt}) and the left inequality in~(\ref{eq:rmax})
hold also when $i = t+1$ taking $r_{t+1} \equiv 0$.
\fi
\end{proposition}

\ifPAGELIMIT
\else
The next proposition determines (up to an additive constant)
the largest degree, $d_t$, of any irreducible factor
of a maximal polynomial $s(x)$
(as well as the smallest degree, $d_{t+1}$, of any
irreducible polynomial that does not divide $s(x)$).
\fi

\begin{proposition}
\label{prop:rmax-d_t-crude}
Using the notation of \Proposition~\ref{prop:unconstrained-lowdegrees},
\begin{equation}
\label{eq:dt1-max}
\floor{\log_q (m/8)} < d_t \le d_{t+1} \le \floor{\log_q m} + 1 .
\end{equation}
\end{proposition}

\ifPAGELIMIT
    We then obtain
\else
We then prove in Section~\ref{sec:lower-upper-bounds}
\fi
the following two bounds.

\begin{theorem}
\label{thm:upper-bound}
For all $m \in \Integers^+$:
\[
\log_2
\Upsilon_m \le \frac{m}{\log_q m}\cdot\left(1 + \O \left(
\frac{\log_q\log_q m}{\log_q m}\right)\right)
.
\]
\end{theorem}

\begin{theorem}
\label{thm:lower-bound}
For all $n \in \Integers^+$:
\[
\log_2 \Upsilon_{n,n} \ge \frac{2n}{\log_q n}
\cdot \left(1-\O\left(\frac{1}{\log_q n}\right)\right)
.
\]
\end{theorem}

\ifPAGELIMIT
\else
The hidden constants in the $\O(\cdot)$ terms in both theorems
are absolute and independent of~$n$, $m$ and~$q$.
\fi
\Theorems~\ref{thm:upper-bound}
and~\ref{thm:lower-bound},
along with
$\Upsilon_{n,n} \le \Upsilon_{2n} \le \Upsilon_{2n+1}$,
imply~(\ref{eq:maxNumDivisors})
and~(\ref{eq:maxNumDivisors(n,n)}).


\emph{Finer characterization of maximal polynomials.}
Our second set of
\ifPAGELIMIT
    results
\else
results, which we prove in
Section~\ref{sec:characterization-unconstrained},
\fi
extends \Proposition~\ref{prop:unconstrained-lowdegrees}.
First, we prove the following estimate for the value of $\rho$.

\begin{proposition}
\label{prop:unconstrained-rmaxexact}
Using the notation of \Proposition~\ref{prop:unconstrained-lowdegrees},
\[
\rho = \frac{\log_q m}{\ln 2} \pm \O \left( \log_q \log_q m \right).
\]
\end{proposition}

Then, we prove the following theorem, which improves
on \Proposition~\ref{prop:unconstrained-lowdegrees}
for large degrees $d_i$.

\begin{theorem}
\label{thm:unconstrained-highdegrees}
Let $s(x) \in \calM_m$ be maximal.
For every $i \in [1:t]$ such that
$d_i \ge \Theta \left( \log_q \log_q m \right)$:
\begin{eqnarray*}
\lefteqn{
\ifPAGELIMIT
    \log_2 \bigl( 1 + (1/(r_i{+}1)) \bigr)
\else
\log_2 \left(  1 + \frac{1}{r_i{+}1} \right)
\fi
\cdot \floor{\log_q m} - \O(1)
} \makebox[10ex]{} \\
&&
\ifIEEE
    {} < d_i \le
\else
{} < \;\; d_i \;\; \le \;\;
\fi
\ifPAGELIMIT
    \log_2 \bigl( 1 + (1/r_i) \bigr)
\else
\log_2 \left(  1 + \frac{1}{r_i} \right)
\fi
\cdot \floor{\log_q m} + \O(1) .
\end{eqnarray*}
\ifPAGELIMIT
    Equivalently:
    $r_i = \floor{ 1\! \left/
        \left(
        2^{(d_i \pm \O(1))/ \lfloor\log_q m\rfloor} - 1 \right)
        \right.}$.
\else
Equivalently,\footnote{
We have made little effort to optimize over the hidden
constants in the $\O(\cdot)$ terms. Our analysis implies that
the expression $\pm \O(1)$ herein has absolute value
at most~$3$. Similarly, the multiplying constant
in the $\O(\log_q \log_q m)$ term in
\Proposition~\ref{prop:unconstrained-rmaxexact} is
only slightly greater than~$3$.}
\[
r_i = \floor{ 1\! \left/
\left(
2^{(d_i \pm \O(1))/\lfloor\log_q m\rfloor} - 1 \right)
\right. }  .
\]
\fi
\end{theorem}

\ifPAGELIMIT
\else
If we substitute $r_i = 2$ in
\Theorem~\ref{thm:unconstrained-highdegrees},
we get that $r_i > 1$ only when
$d_i/\log_q m  < \log_2 (3/2) + o_m(1) \approx 0.585$.
Combining this with
\Proposition~\ref{prop:rmax-d_t-crude},
we conclude that for a given~$q$ and $m \rightarrow \infty$,
all but a vanishing fraction of the multiplicities in
$\hist(s)$ are~$1$.
\fi


\emph{Characterization of $n$-maximal polynomials.}
Our third set of
\ifPAGELIMIT
    results
\else
results, which we prove in
Section~\ref{sec:characterization-constrained},
\fi
addresses the second part of Problem~\ref{prob:constrained}
and provides a characterization of an $n$-maximal polynomial.
\ifPAGELIMIT
\else
We introduce some notation.
\fi

For $n \in \Integers^+$ and
$s(x) = \prod_{i=1}^t p_i(x)^{r_i} \in \calP_{2n}$,
let $r_0 = 2n - \deg s$ and write
\ifPAGELIMIT
    $\hist_n(s) = (r_i)_{i=0}^t$.
\else
$\hist_n(s) = (r_0 \; \hist(s)) = (r_i)_{i=0}^t$.
\fi
Also, define
\[
\rho_n(s) = \max \{ r_0, \rho(s) \}
= \max_{i \in \Integers_{\ge 0} \,:\, d_i = 1} r_i ,
\]
where $d_0 \equiv 1$.
\Proposition~\ref{prop:nohole}
through \Theorem~\ref{thm:unconstrained-highdegrees}
hold also for $n$-maximal polynomials,
with~$m$, $\hist(s)$, and $\rho(s)$
\ifPAGELIMIT
    replaced by $2n$, $\hist_n(s)$, and $\rho_n(s)$, respectively.
\else
therein replaced by $2n$, $\hist_n(s)$, and $\rho_n(s)$,
respectively, and the index~$i$ also allowed to be~$0$.
\fi
In particular, the counterpart of
\Proposition~\ref{prop:unconstrained-lowdegrees}
reads as follows.

\begin{proposition}
\label{prop:constrained-lowdegrees}
Let $s(x) \in \calP_{2n}$ be $n$-maximal
and let $\hist_n(s) = (r_i)_{i=0}^t$ and $\rho_n = \rho_n(s)$.
For every $i \in [0:t]$:
\begin{equation}
\label{eq:constrained-lowdegrees}
\frac{\rho_n + 1}{r_i + 2} \le d_i < \frac{\rho_n + 1}{r_i} .
\end{equation}
\ifPAGELIMIT
    Equivalently:
    $r_i \in \left\{\floor{\rho_n/d_i},\
    \floor{\rho_n/d_i}-1 \right\}$.
\else
Equivalently,
\begin{equation}
\label{eq:ri-formula}
r_i \in \left\{\floor{\frac{\rho_n}{d_i}},\
\floor{\frac{\rho_n}{d_i}}-1 \right\} .
\end{equation}
Moreover, (\ref{eq:ri-formula})
and the left inequality in~(\ref{eq:constrained-lowdegrees})
hold also when $i = t+1$ taking $r_{t+1} \equiv 0$.
\fi
\end{proposition}

\ifPAGELIMIT
    It
\else
Unlike~(\ref{eq:factorization}),
we do not have a simple expression for $\ndiv_n(s)$.
Therefore, our results for $n$-maximal polynomials
(such as \Proposition~\ref{prop:constrained-lowdegrees})
require more intricate proofs than those for maximal polynomials.
Moreover, it
\fi
follows from the $n$-maximal counterparts of
\Propositions~\ref{prop:unconstrained-lowdegrees}
and~\ref{prop:unconstrained-rmaxexact}
that $r_0 = \Theta(\log_q n)$;
namely, any $n$-maximal polynomial $s(x) \in \calP_{2n}$
\ifPAGELIMIT
    has degree $2n - \Theta(\log_q n) \; (< 2n)$.
    In contrast, recall that the maximum in~(\ref{eq:unconstrained})
    is attained by a polynomial $s(x)$ of degree \emph{exactly}~$m$.
\else
has degree $2n - \Theta(\log_q n) < 2n$.
Thus, while the maximum in~(\ref{eq:unconstrained})
is attained by a polynomial $s(x)$ of degree exactly~$m$,
the maximum in~(\ref{eq:constrained}) is attained
by a polynomial of degree strictly less than $n + n' = 2n$.
\fi

\emph{Average-case analysis.}
In our fourth set of results,
\ifPAGELIMIT
\else
which will be the subject of
Section~\ref{sec:average-case},
\fi
we consider the probabilistic counterparts of
Problems~\ref{prob:unconstrained} and~\ref{prob:constrained}.
In the case of ordinary factorizations,
given $m \in \Integers^+$, we take
the sample space to be $\calM_m$,
assume a uniform distribution over $\calM_m$,
and define a random variable
\ifPAGELIMIT
    $\random_m$
\else
$\random_m = \random_m(q)$
\fi
over $s(x) \in \calM_m$ by
$\random_m : s \mapsto \ndiv(s)$.
\ifPAGELIMIT
\else
We prove the following theorem.
\fi

\begin{theorem}
\label{thm:expectation-variance-unconstrained}
\[
\Expected \left\{ \random_m \right\} = m+1
\ifIEEE
    \quad \textrm{and} \quad
\else
\quad\quad \textrm{and} \quad\quad
\fi
\Variance \left\{ \random_m \right\} = \frac{q-1}{q}  \binom{m+1}{3} .
\]
\end{theorem}

Using the well-known Markov and Chebyshev
\ifPAGELIMIT
    inequalities
\else
inequalities~\cite[p.~127]{Gallager}
\fi
we get that for every $\varepsilon > 0$,
\[
\Prob \left\{ \random_m \ge m^{1+\varepsilon} \right\} \le
\O \bigl( m^{- \max \{ \varepsilon, 2\varepsilon - 1 \}} \bigr) .
\]
In particular, the probability of $\random_m$
being super-linear in $m$ tends to~$0$ as $m \rightarrow \infty$.
Through a different approach, which uses the Chernoff bound,
we are also able to prove the following result,
which implies that the median of $\random_m$ is sub-linear in~$m$.

\begin{proposition}
\label{prop:chernoff-tail}
For any (fixed) $\varepsilon > 0$,
\[
\Prob \left\{ \random_m \ge m^{\varepsilon + \ln 2} \right\}
\le \O \bigl( m^{-\kappa(\varepsilon)} \bigr)
,
\]
where $\kappa(\varepsilon) > 0$.
\end{proposition}

\ifPAGELIMIT
\else
The proof of the proposition can be found
\ifJCTA
    in \ref{sec:chernoff-tail}.
\else
    in Appendix~\ref{sec:chernoff-tail}.
\fi
\fi

In the case of $(n,n)$-factorizations,
we consider a different probability model,
which fits better the coding application that was mentioned
in Section~\ref{sec:introduction}, namely,
the list decoding of the rank-metric code of~\cite{Roth18},
assuming error arrays that are uniformly distributed conditioned
on having rank $1$.
Accordingly,
given $n \in \Integers^+$, the sample space is defined to be
$\calP_n^2 = \calP_n \times \calP_n$,
over which we assume a uniform distribution.
We define a random variable
\ifPAGELIMIT
    $\random_{n,n}$
\else
$\random_{n,n} = \random_{n,n}(q)$
\fi
over $(u,v) \in \calP_n^2$ by
$\random_{n,n} : (u,v) \mapsto \ndiv_n(u \cdot v)$
(i.e., the number of $(n,n)$-factorizations of the product $u \cdot v$).
\ifPAGELIMIT
\else
We prove the following theorem.
\fi

\begin{theorem}
\label{thm:expectation-variance-constrained}
\[
\Expected \left\{ \random_{n,n} \right\} = (n+1)(1+\O(1/q))
\ifIEEE
    \;\;\; \textrm{and} \;\;\;
\else
\quad\quad \textrm{and} \quad\quad
\fi
\Variance \left\{ \random_{n,n} \right\} = \O(n^4)
\ifPAGELIMIT
   .
   \]
\else
,
\]
where the hidden constants in the $\O(\cdot)$ terms are
absolute and independent of~$q$ and~$n$.
\fi
\end{theorem}

Thus, $\random_{n,n}$, too, takes super-linear values in~$n$
with vanishing probability as $n \rightarrow \infty$.
\ifPAGELIMIT

    Due to space limitations,
    many proofs in this abstract are either sketched or omitted.
    The full text can be found in~\cite{BR}.
\else
We also show that the $\O(n^4)$ expression for
$\Variance \left\{ \random_{n,n} \right\}$
in \Theorem~\ref{thm:expectation-variance-constrained}
can be tightened to $\Theta(n^4)$, at least for $q \ge 9$.
\fi

\section{Bounds on $\Upsilon_m$ and $\Upsilon_{n,n}$}
\label{sec:lower-upper-bounds}

\ifPAGELIMIT
\else
This section is devoted to proving
\Proposition~\ref{prop:nohole} through \Theorem~\ref{thm:lower-bound}.
\fi

For $d \in \Integers^+$, let
$\I(d) = \I(d,q)$ be the number of monic irreducible polynomials
of degree~$d$ over~$\F$.
\ifPAGELIMIT
    It follows from~\cite[\Theorem~3.25]{LN}
\else
This number is given by the expression
\[
\I(d)
= \frac{1}{d} \sum_{\ell \in \Integers^+ \,:\, \ell \,|\, d}
\mu(\ell) \cdot q^{d/\ell} ,
\]
where~$\mu(\cdot)$ is the Moebius function~\cite[\Theorem~3.25]{LN}.
It follows
\fi
that for any $d \in \Integers^+$,
\begin{equation}
\label{eq:bounds-on-I(d)}
\frac{1}{d} \left( q^d - 2q^{\floor{d/2}} \right)
< \I(d)
\le \frac{q^d}{d} ,
\end{equation}
and by induction on~$d$ we readily get:
\begin{equation}
\label{eq:J(d)-estimation}
\sum_{\ell=1}^d \I(\ell)
\le \sum_{\ell=1}^d \frac{q^\ell}{\ell} < \frac{4q^d}{d+1} .
\end{equation}

We proceed to proving
\ifPAGELIMIT
    \Proposition~\ref{prop:nohole} through
    \Theorem~\ref{thm:lower-bound}.
    Many of the proofs in this work
    follow a similar pattern: we assume
\else
\Propositions~\ref{prop:nohole}
and~\ref{prop:unconstrained-lowdegrees}.
Many of the proofs in this work
will follow a similar pattern: we will assume
\fi
that a polynomial
$s \in \calP_m$ does \emph{not} satisfy the property to be proved,
and we construct from~$s$
a polynomial $\tis \in \calP_m$ for which
$\ndiv(\tis) > \ndiv(s)$, thereby showing
that~$s$ cannot be maximal.
\ifPAGELIMIT
    Due to space limitations, in most cases we will only indicate
    what~$\tis$ is; the full details can be found in~\cite{BR}.
\fi

\begin{proof}[Proof of \Proposition~\ref{prop:nohole}]
Given $d_j < d_i$, assume that
$s(x) \in \calP_m$ is such that $r_i \ge r_j + 1$,
and let $p_k(x) \in \calM_1$ where $k \ne j$.
The polynomial
\[
\tis(x) = s(x) \cdot p_k(x) \cdot p_j(x) / p_i(x)
\]
is in $\calP_m$ and satisfies
\[
\ndiv(\tis)
\stackrel{(\ref{eq:factorization})}{=}
\frac{r_k+2}{r_k+1} \cdot
\frac{r_j+2}{r_j+1} \cdot
\frac{r_i}{r_i+1}
\cdot \ndiv(s)
\ge
\frac{r_k+2}{r_k+1}
\cdot \ndiv(s)
> \ndiv(s) .
\]
Thus, $s$ cannot be maximal.
\end{proof}


\ifPAGELIMIT
    \begin{proof}[Proof sketch of
    \Proposition~\ref{prop:unconstrained-lowdegrees}]
    Assuming that $s(x)$ does not satisfy
    the left inequality in~(\ref{eq:rmax}), define
    \[
    \tis(x) = s(x) \cdot p_i(x) / p_k(x)^{d_i} ,
    \]
    where $p_k(x) \in \calM_1$ is such that $\rho = r_k$.
    Then $\ndiv(\tis) > \ndiv(s)$.
    The right inequality in~(\ref{eq:rmax}) is proved
    in a similar manner, taking
    \[
    \tis(x) = s(x)
    \cdot p_k(x)^{\floor{d_i/2}}
    \cdot p_\ell(x)^{\ceil{d_i/2}} / p_i(x) ,
    \]
    where $p_k(x)$ and $p_\ell(x)$ are distinct in $\calM_1$.
\end{proof}
\else
\begin{proof}[Proof of \Proposition~\ref{prop:unconstrained-lowdegrees}]
Starting with the left inequality in~(\ref{eq:rmax}),
let $p_k(x) \in \calM_1$ be such that $\rho = r_k$
and suppose that $s(x) \in \calM_m$ is such that
$r_i < (\rho + 1)/d_i - 2$
(in particular, we must have $i \ne k$ and $d_i \le \rho$);
this implies that
\begin{equation}
\label{eq:rmax3}
\frac{\rho - d_i + 1}{\rho + 1} \cdot \frac{r_i + 2}{r_i + 1} > 1 .
\end{equation}
Define
\[
\tis(x) = s(x) \cdot p_i(x) / p_k(x)^{d_i}
\]
(which is a proper polynomial since $d_i \le \rho = r_k$).
We have $\deg \tis = \deg s = m$ and
\[
\ndiv(\tis)
\stackrel{(\ref{eq:factorization})}{=}
\frac{\rho - d_i + 1}{\rho + 1}
\cdot \frac{r_i + 2}{r_i + 1} \cdot \ndiv(s)
\stackrel{(\ref{eq:rmax3})}{>} \ndiv(s) .
\]
Notice that the proof holds also when $i = t+1$.

Turning to the right inequality in~(\ref{eq:rmax}),
suppose that $s(x) \in \calM_m$ is such that $r_i \ge (\rho + 1)/d_i$
(in particular, we must have $d_i \ge 2$); this implies that
\begin{equation}
\label{eq:rmax1}
\frac{\rho + d_i + 1}{\rho + 1} \cdot \frac{r_i}{r_i + 1} \ge 1 ,
\end{equation}
with equality \ifandonlyif\ $r_i = (\rho + 1)/d_i$.
Consider the polynomial
\[
\tis(x) = s(x) \cdot p_k(x)^{d_i} / p_i(x) .
\]
We have $\deg \tis = \deg s = m$ and
\begin{equation}
\label{eq:rmax2}
\ndiv(\tis)
\stackrel{(\ref{eq:factorization})}{=}
\frac{\rho + d_i + 1}{\rho + 1} \cdot \frac{r_i}{r_i + 1}
\cdot \ndiv(s)
\stackrel{(\ref{eq:rmax1})}{\ge} \ndiv(s) ,
\end{equation}
with equality \ifandonlyif\ $r_i = (\rho + 1)/d_i$.
Thus, if the inequality in~(\ref{eq:rmax2}) is strict, we are done.
Otherwise,
letting $p_j(x) \in \calM_1$ be other than $p_k(x)$,
we have $\tir_j = \mult_{p_j}(\tis) = r_j \le r_k = \rho$.
Therefore,
\[
\tilde{\rho} = \rho(\tis) = r_k + d_i \ge r_j + 2 = \tir_j + 2
\]
and, so,
\[
\frac{\tilde{\rho} + 1}{\tir_j + 2} > 1 = d_j .
\]
This means that~$\tis$ (and, therefore, $s$) cannot be maximal,
since it violates the left inequality in~(\ref{eq:rmax}).
\end{proof}
\fi

\ifPAGELIMIT
\else
\begin{remark}
\label{rem:AMGM}
The reciprocal relation between $d_i$ and $r_i$
in~(\ref{eq:rmax}) is somewhat expected.
Given~$m$ and conditioning on the value of~$t$,
the maximization of the expression~(\ref{eq:factorization})
over the \emph{real} vectors $(r_i)_{i=1}^t$,
subject to the linear constraint $\sum_{i=1}^t r_i d_i = m$, yields
\[
r_i = \frac{c}{d_i} - 1 ,
\]
where
$c = (1/t) \bigl( 2n + \sum_{i=1}^t d_i \bigr)$.\qed
\end{remark}

We will use the next lemma in upcoming proofs.
\fi

\begin{lemma}
\label{lem:rmax-d_t-crude}
Using the notation of \Proposition~\ref{prop:unconstrained-lowdegrees},
\begin{equation}
\label{eq:dt<rho<2dt}
d_t \le \rho  \le 2 d_{t+1} - 1 .
\end{equation}
\end{lemma}

\begin{proof}
Substituting $i = t$ (\respectively, $i = t+1$)
in \Proposition~\ref{prop:unconstrained-lowdegrees}
yields the left (\respectively, right) inequality.
\end{proof}

\begin{proof}[Proof of \Proposition~\ref{prop:rmax-d_t-crude}]
The following chain of inequalities imply
the leftmost inequality in~(\ref{eq:dt1-max}):
\begin{eqnarray*}
m = \deg s & = &
\sum_{i=1}^t r_i d_i
\stackrel{(\ref{eq:rmax})}{\le} t(\rho+1)
\stackrel{(\ref{eq:dt<rho<2dt})}{\le} 2 d_{t+1} \cdot t \\
& \le &
2 d_{t+1} \cdot \sum_{\ell=1}^{d_t} \I(\ell)
\stackrel{(\ref{eq:J(d)-estimation})}{<}
8 q^{d_t} .
\end{eqnarray*}
As for the rightmost inequality in~(\ref{eq:dt1-max}),
we recall from~\cite[Corollary~3.21]{LN} that
$q^d = \sum_{\ell|d} \ell \cdot \I(\ell)$;
hence, by \Proposition~\ref{prop:nohole},
\[
m = \deg s \ge
\sum_{\ell=1}^{d_{t+1} - 1} \ell \cdot \I(\ell)
\ge q^{d_{t+1}-1} .
\]
\end{proof}

\begin{proof}[Proof of \Theorem~\ref{thm:upper-bound}]
Let $s(x) \in \calP_m$ be maximal,
\ifPAGELIMIT
    let $\varepsilon \in(0,1)$,
\else
let $\varepsilon = \varepsilon(m) \in(0,1)$ (to be determined shortly),
\fi
and consider first all the irreducible factors of $s(x)$ of
degree at most ${}\Delta = \floor{ (1-\varepsilon)\log_q m}$.
\ifPAGELIMIT
    By
    \Proposition~\ref{prop:unconstrained-lowdegrees}
    and Lemma~\ref{lem:rmax-d_t-crude},
    their total number, $w_1$ (counting multiplicities),
\else
By \Proposition~\ref{prop:unconstrained-lowdegrees}
and  Lemma~\ref{lem:rmax-d_t-crude},
the total number, $w_1$, of such factors,
\emph{counting multiplicities},
\fi
is bounded from above by
\begin{eqnarray*}
w_1 & = & \sum_{i\,:\,d_i \le \Delta} r_i
\le
\sum_{d=1}^\Delta
\frac{2 d_{t+1}}{d} \cdot \I(d)
\le 2 d_{t+1} \cdot \sum_{d=1}^\Delta \I(d) \\
& \stackrel{(\ref{eq:J(d)-estimation})}{\le} &
2 d_{t+1} \cdot
\frac{4 q^\Delta}{\Delta+1}
\stackrel{(\ref{eq:dt1-max})}{=}
\O \left( m^{1-\varepsilon} \right) .
\end{eqnarray*}
Selecting $\varepsilon = 2 (\log_q \log_q m)/\log_q m$,
we readily get:
\[
w_1 = \O \left( m^{1-\varepsilon} \right)
= \O \left( m / \log_q^2 m \right) .
\]
Turning to the irreducible factors of $s(x)$ whose degrees
exceed~$\Delta$, their total number, $w_2$
(counting multiplicities), is bounded from above by
\begin{eqnarray*}
w_2 & \le &
\frac{m}{\Delta+1} <
\frac{m}{(1 - \varepsilon)\log_q m} \\
& = &
\frac{m}{\log_q m}
\cdot
\left(1 + \O \left( \frac{\log_q\log_q m}{\log_q m} \right) \right).
\end{eqnarray*}
We conclude that
\[
\log_2 \ndiv(s) \le w_1 + w_2
\le
\frac{m}{\log_q m}
\cdot
\left(1 + \O \left( \frac{\log_q\log_q m}{\log_q m} \right) \right) .
\]
\end{proof}

\begin{proof}[Proof of \Theorem~\ref{thm:lower-bound}]
Let~$d$ be the smallest integer such that
$d \cdot \I(d) \ge 2n$;
by~(\ref{eq:bounds-on-I(d)}) we have
$d \in \left\{ \ceil{\log_q n}{+}1,
\ceil{\log_q n}{+}2 \right\}$.
Let $w = \floor{n/d}$,
and let $s(x)$ be a product of~$2w$ distinct
monic irreducible polynomials
of degree~$d$. Such a polynomial has degree $\le 2n$ and
$\binom{2w}{w}$ distinct $(n,n)$-factorizations.
We have:
\[
\Upsilon_{n,n} \ge \ndiv_n(s) = \binom{2w}{w}
= 2^{(2n/\log_q n) \cdot\left(1-\O\left(1/\log_q n\right)\right)},
\]
where the last equality follows from
$w = (n/\log_q n)(1 - \O(1/\log_q n))$
and known approximations of the binomial
coefficients~\cite[p.~309, Eq.~(16)]{MS}.
\end{proof}

\section{Characterization of maximal polynomials}
\label{sec:characterization-unconstrained}

\ifPAGELIMIT
    \Proposition~\ref{prop:unconstrained-rmaxexact}
    and \Theorem~\ref{thm:unconstrained-highdegrees}
    are proved using the next two lemmas.
\else
In this section, we prove
\Proposition~\ref{prop:unconstrained-rmaxexact}
and \Theorem~\ref{thm:unconstrained-highdegrees}.
The proof technique bears resemblance to the proofs
in~\cite[\S4]{AE} on the structural properties
of highly-composite integers, namely,
integers that have more divisors than any smaller integer.
\fi
Hereafter,
we let $\delta_q(m)$ be the smallest positive integer~$\delta$
such that $\I(d) > \floor{\log_q m} + 1$ for every $d \ge \delta$.
By~(\ref{eq:bounds-on-I(d)}), it follows that
$\delta_q(m)=\log_q \log_q (q \, m)  +o \left( \log_q \log_q m \right)$.

\begin{lemma}
\label{lem:diupper}
Let $s(x) \in \calM_m$ be maximal and let $i \in [1:t]$.
\begin{list}{}{\settowidth{\labelwidth}{\textit{(ii)}}}
\item[(a)]
If $d_i \ge \delta_q(m)$ and $r_i > 1$ then
\begin{equation}
\label{eq:diupper1}
d_i \le \log_2 %
\ifPAGELIMIT
    \bigl(  r_i/(r_i{-}1) \bigr)
\else
    \left(  \frac{r_i}{r_i - 1} \right)
\fi
\cdot (d_t + 1) .
\end{equation}
\item[(b)]
If $d_i \ge \delta_q(m) + 1$ then
\[
d_i \le \log_2 %
\ifPAGELIMIT
    \bigl(  (r_i{+}1)/r_i \bigr)
\else
    \left(  \frac{r_i + 1}{r_i} \right)
\fi
\cdot (d_t + 1) + 1 .
\]
\end{list}
\end{lemma}

\begin{proof}
\emph{(a)}
Let~$\Set$ be a set of $d_t + 1$ indexes~$j$ for which $d_j = d_i$;
  from $d_i \ge \delta_q(m)$
and \Proposition~\ref{prop:rmax-d_t-crude} we have
$\I(d_i) > \floor{\log_q m} + 1 \ge d_t$ and, so, such a set
indeed exists. Also, let~$\Setalt$ be a set of $d_i$ 
indexes~$k$ for which
\ifPAGELIMIT
    $d_k = d_t + 1$.
\else
$d_k = d_t + 1$; such a set exists too. Note that
\fi
\Proposition~\ref{prop:unconstrained-lowdegrees} implies that
$r_j \ge r_i - 1 \; (> 0)$ when $j \in \Set$.
\ifPAGELIMIT
\else
Since $r_k = 0$ when $k \in \Setalt$,
it follows that $\Set \cap \Setalt = \emptyset$.

\fi
Define the polynomial
\[
\ifPAGELIMIT
    \textstyle
\fi
\tis(x) = s(x) \cdot \Bigl( \prod_{k \in \Setalt} p_k(x) \Bigr)
\Bigm/ \prod_{j \in \Set} p_j(x) .
\]
We have:
\begin{equation}
\label{eq:diupper2}
\frac{\ndiv(\tis)}{\ndiv(s)}
\stackrel{(\ref{eq:factorization})}{=}
2^{|\Setalt|} \cdot \prod_{j \in \Set} \frac{r_j}{r_j + 1}
\ge
2^{d_i} \cdot
\left( \frac{r_i - 1}{r_i} \right)^{d_t+1} .
\end{equation}
Now, $\deg \tis = \deg s = m$ and, so,
$\ndiv(\tis)/\ndiv(s) \le 1$ (since $s$ is maximal).
The result follows from~(\ref{eq:diupper2}) by taking logarithms.

\emph{(b)}
The proof is similar to part~(a), except that $d_i$ is replaced
by $d_i - 1 \; (> 0)$:
now~$\Set$ is a set of $d_t + 1$ indexes~$j$ for which
$d_j = d_i - 1$, and~$\Setalt$ is a set of $d_i - 1$ 
indexes~$k$ for which
\ifPAGELIMIT
    $d_k = d_t + 1$.
\else
$d_k = d_t + 1$ (both sets exist when $d_i - 1 \ge \delta_q(m)$).
\Proposition~\ref{prop:nohole} implies that
$r_j \ge r_i \; (> 0)$ when $j \in \Set$ and, thus,
$\Set \cap \Setalt = \emptyset$.
Re-defining $\tis(x)$ with these sets $\Set$ and $\Setalt$,
we get that~(\ref{eq:diupper2})
holds, with $d_i$ and $r_i$ therein replaced by $d_i - 1$ and
$r_i + 1$, respectively.
\fi
\end{proof}

\begin{lemma}
\label{lem:dilower}
Let $s(x) \in \calM_m$ be maximal and let $i \in [1:t]$.
\begin{list}{}{\settowidth{\labelwidth}{\textit{(ii)}}}
\item[(a)]
If $d_i \ge \delta_q(m)$ then
\[
d_i \ge \log_2 %
\ifPAGELIMIT
    \bigl( (r_i{+}3)/(r_i{+}2) \bigr)
\else
    \left( \frac{r_i + 3}{r_i + 2} \right)
\fi
\cdot (d_{t+1} - 1) .
\]
\item[(b)]
If $d_i \ge \delta_q(m) - 1$ then
\[
d_i \ge \log_2 %
\ifPAGELIMIT
    \bigl( (r_i{+}2)/(r_i{+}1) \bigr)
\else
    \left(  \frac{r_i + 2}{r_i + 1} \right)
\fi
\cdot (d_{t+1} - 1) - 1 .
\]
\end{list}
\end{lemma}

\ifPAGELIMIT
    \begin{proof}[Proof sketch]
    Here we take~$\Set$ to be a set of $d_{t+1} - 1$ indexes~$j$
    for which $d_j = d_i$ (\respectively, $d_j = d_i + 1$
    for part~(b))
    and~$\Setalt$ to be set of $d_i$ (\respectively, $d_i + 1$)
    indexes~$k$ for which $d_k = d_{t+1} - 1$.
    The polynomial~$\tis$ is defined by
    \[
    \textstyle
    \tis(x) = s(x) \cdot \Bigl( \prod_{j \in \Set} p_j(x) \Bigr)
    \Bigm/ \prod_{k \in \Setalt} p_k(x) .
    \]
    \end{proof}
\else
\begin{proof}
\emph{(a)}
The claim trivially holds when $d_i \ge d_{t+1} - 1$, so we assume
hereafter in the proof that $d_i < d_{t+1} - 1$.
Let~$\Set$ be a set of $d_{t+1} - 1$ indexes~$j$
for which $d_j = d_i$
and let~$\Setalt$ be a set of $d_i$ 
indexes~$k$ for which $d_k = d_{t+1} - 1$.
Note that $\Set \cap \Setalt = \emptyset$
(since $d_i < d_{t+1} - 1$)
and $r_k > 0$ when $k \in \Setalt$.
Also, $r_j \le r_i + 1$ when $j \in \Set$
(by \Proposition~\ref{prop:unconstrained-lowdegrees}).

Define
\[
\tis(x) = s(x) \cdot \Bigl( \prod_{j \in \Set} p_j(x) \Bigr)
\Bigm/ \prod_{k \in \Setalt} p_k(x) .
\]
We have:
\begin{equation}
\label{eq:dilower}
\frac{\ndiv(\tis)}{\ndiv(s)}
\stackrel{(\ref{eq:factorization})}{=}
\prod_{k \in \Setalt}
\frac{r_k}{r_k + 1}
\cdot
\prod_{j \in \Set}
\frac{r_j + 2}{r_j + 1}
\ge 2^{-d_i} \cdot
\left( \frac{r_i + 3}{r_i + 2} \right)^{d_{t+1}-1} .
\end{equation}
We now proceed as in Lemma~\ref{lem:diupper}(a):
$\deg \tis = \deg s = m$ implies that
$\ndiv(\tis)/\ndiv(s) \le 1$,
and the result follows by taking logarithms.

\emph{(b)}
The claim is trivial when $d_i \ge d_{t+1} - 2$;
for smaller $d_i$ we modify the proof of part~(a) as follows.
We take~$\Set$ to be a set of $d_{t+1} - 1$ indexes~$j$
for which $d_j = d_i + 1$
and~$\Setalt$ to be a set of $d_i + 1$ 
indexes~$k$ for which $d_k = d_{t+1} - 1$.
We again have $\Set \cap \Setalt = \emptyset$
(since $d_i < d_{t+1} - 2$)
and $r_k > 0$ when $k \in \Setalt$.
Also, $r_j \le r_i$ when $j \in \Set$
(by \Proposition~\ref{prop:unconstrained-lowdegrees}).
Re-defining $\tis(x)$, we get that~(\ref{eq:dilower})
holds, with $d_i$ and $r_i$ therein replaced by $d_i + 1$ and
$r_i - 1$, respectively.
\end{proof}

In each of the previous two lemmas,
part~(a) is stronger when $d_i$ is small (and $r_i$ is large),
whereas part~(b) is more effective for large $d_i$.
\fi

\begin{proof}[Proof of \Proposition~\ref{prop:unconstrained-rmaxexact}]
Let $i \in [1\,{:}\,t]$ be such that
$d_i = \delta_q(m)\, (= \O(\log_q \! \log_q {\scriptscriptstyle\!} m))$.
By Lemma~\ref{lem:diupper}(a)
\ifPAGELIMIT
\else
and the inequality $e^z \ge 1 + z$
\fi
we have
\begin{equation}
\label{eq:rmaxexact1}
1 + \frac{1}{r_i - 1} \ge 2^{d_i/(d_t + 1)}
\ge 1 + \frac{d_i \ln 2}{d_t + 1}
\end{equation}
and, so, along with \Proposition~\ref{prop:unconstrained-lowdegrees}
we obtain:
\[
\floor{\frac{\rho}{d_i}} - 1 \le r_i
\le \frac{d_t + 1}{d_i \ln 2} + 1 .
\]
Hence,
\begin{equation}
\label{eq:rmaxexact2}
\rho < \frac{d_t + 1}{\ln 2} + 3 d_i
= \frac{\log_q m}{\ln 2} + \O \left( \log_q \log_q m \right)
\ifPAGELIMIT
    .
    \end{equation}
    The proof of the lower bound on~$\rho$ is similar, except
    that we use Lemma~\ref{lem:dilower}(a) instead.
\else
,
\end{equation}
where the last step follows from \Proposition~\ref{prop:rmax-d_t-crude}.

Turning to bounding~$\rho$ from below,
by Lemma~\ref{lem:dilower}(a)
and the inequality $e^z < 1/(1 - z) = 1 + (1/z - 1)^{-1}$
over $z \in (0,1)$ we get:
\[
1 + \frac{1}{r_i + 2} \le 2^{d_i/(d_{t+1} - 1)}
< 1 + \left( \frac{d_{t+1} - 1}{d_i \ln 2} - 1 \right)^{-1} .
\]
Combining with \Proposition~\ref{prop:unconstrained-lowdegrees} yields:
\[
\floor{\frac{\rho}{d_i}} \ge r_i > \frac{d_{t+1} - 1}{d_i \ln 2} - 3 ,
\]
namely,
\[
\rho > \frac{d_{t+1} - 1}{\ln 2} - 3 d_i
= \frac{\log_q m}{\ln 2} - \O \left( \log_q \log_q m \right) .
\]
\fi
\end{proof}

\begin{proof}[Proof of \Theorem~\ref{thm:unconstrained-highdegrees}]
Combine Lemmas~\ref{lem:diupper}(b)
and~\ref{lem:dilower}(b) with
\Proposition~\ref{prop:rmax-d_t-crude}.
\end{proof}

\section{Characterization of $n$-maximal polynomials}
\label{sec:characterization-constrained}

Given $n \in \Integers^+$ and $s(x) \in \calP_{2n}$,
\ifPAGELIMIT
\else
for convenience
\fi
we extend the degree of $s(x)$ to $2n$
by introducing a slack variable~$y$ and defining
\begin{equation}
\label{eq:s(x,y)}
s(x,y) = y^{r_0} \cdot s(x) ,
\end{equation}
where $r_0 = 2n - \deg s(x)$.
Accordingly, we introduce the following notation:
\[
\Comp_m = \Comp_m(q)
= \left\{ y^{m-\deg u} \cdot u(x) \,:\, u(x) \in \calP_m \right\} .
\]
Given $b(x,y) \in \Comp_m$,
we denote by $\calD_k(b)$
the set of divisors of $b(x,y)$ in $\Comp_k$.
Thus $s(x,y) \in \Comp_{2n}$,
and there is a one-to-one correspondence between
the $(n,n)$-factorizations
$(u(x),v(x)) \in \calP_n^2$ of $s(x)$
and divisors $u(x,y) \in \calD_n(s(x,y))$.
In particular, $\ndiv_n(s) = |\calD_n(s)|$.

Given a polynomial
$s(x,y) \in \Comp_{2n}$, fix a factorization
\begin{equation}
\label{eq:factorization-a}
s(x,y) = a(x) \cdot b(x,y) ,
\end{equation}
where $\gcd(a,b) = 1$ and $b(x,y) \in \Comp_\degb$,
for some $\degb \in [r_0:2n]$
(we will determine~$a$ and~$b$ later).
For every $k \in [\degb{-}n:n]$ let
\begin{equation}
\label{eq:A}
\calA_k = \calA_k(n,a) = \left\{ f \in \calM_{n-k}
\,:\, f \,|\, a \right\} .
\end{equation}
We have:
\[
\calD_n(s)
=
\bigcupdot_{k \in [0:\degb]}
\bigl\{ f \cdot \eta
\,:\, (f,\eta) \in \calA_k \times \calD_k(b) \bigr\}
\]
and, so,
\begin{equation}
\label{eq:partitionA}
\ifPAGELIMIT
    \textstyle
\fi
\ndiv_n(s)
= |\calD_n(s)|
= \sum_{k\in [0:\degb]} |\calA_k| \cdot |\calD_k(b)| .
\end{equation}
\ifPAGELIMIT
\else
The decomposition~(\ref{eq:partitionA}) will be used
in several proofs below.

\subsection{Proof of \Proposition~\ref{prop:nohole}
for the $n$-maximal case}
\label{sec:no-zero-holes}

In this section we prove the following proposition,
which is the counterpart of \Proposition~\ref{prop:nohole}
for $n$-maximal polynomials.

\begin{proposition}
\label{prop:no-zero-holes}
Let $s(x) \in \calP_{2n}$ be $n$-maximal
and let $\hist_n(s) = (r_i)_{i=0}^t$.
For any $i, j \in [0:t]$,
if $d_i > d_j$ then $r_i \le r_j$.
\end{proposition}

Fix a polynomial $s(x) = \prod_{i=1}^t {p_i(x)}^{r_i} \in \calP_{2n}$
and let $s(x,y) = y^{r_0} \cdot s(x)$ be as in~(\ref{eq:s(x,y)}).
\Withoutlossofgenerality\ assume that
$\rho_n = \rho_n(s) = r_0$
(otherwise, if, say $\rho_n(s) = r_1$,
we could switch the roles of~$y$ and $p_1(x)$ in
the upcoming analysis).

We make a running assumption that
there exist $i > j$ in $[0:t]$ such that
$d_i > d_j$ and $r_i > r_j$ (since $\rho_n = r_0$
we can assume that $j > 0$); we show that~$s$ cannot be $n$-maximal
by exhibiting a polynomial $\tis$ such that
$\ndiv_n(\tis) > \ndiv_n(s)$.
\Withoutlossofgenerality\ we further assume that
the difference $i - j$ is
the smallest for which $d_i > d_j$ and $r_i > r_j$,
in which case $d_j = d_i - 1$.

Assume the factorization~(\ref{eq:factorization-a}), where
\[
b(x,y) =  y^{r_0} \cdot {p_j(x)}^{r_j} \cdot p_i(x)^{r_i}
\]
(and, therefore, $\gcd(a,b) = 1$),
\[
\degb = \deg b(x,y) = r_0 + r_j d_j + r_i d_i ,
\]
and $\deg a(x) = 2n - \degb$. Also, let
\begin{equation}
\label{eq:bubblesort}
\tis(x) = s(x) \cdot p_j(x) / p_i(x)
\end{equation}
and write
$\tis(x,y) = y^{\tir_0} \cdot \tis(x) = a(x) \cdot \tib(x,y)$,
where
\begin{eqnarray*}
\tib(x,y)
& = & y^{\tir_0} \cdot {p_j(x)}^{\tir_j} \cdot {p_i(x)}^{\tir_i} \\
& = &
y^{r_0+1} \cdot {p_j(x)}^{r_j+1} \cdot {p_i(x)}^{r_i-1}
\end{eqnarray*}
(and $\gcd(a,\tib) = 1$);
namely, the multiplicities of~$y$ and $p_j(x)$ increase by~$1$
while the multiplicity of $p_i(x)$ decreases by~$1$.
We have
$\deg \tib(x,y) = \tir_0 + \tir_j d_j + \tir_i d_i = \degb$
and, so, $\deg \tis(x,y) = \degb + \deg a = \deg s(x,y) = 2n$.
Rewriting~(\ref{eq:partitionA}) for~$\tis$ we get:
\begin{equation}
\label{eq:partitionB}
\ndiv_n(\tis)
= |\calD_n(\tis)|
= \sum_{k \in [0:\degb]}
|\calA_k| \cdot |\calD_k(\tib)| .
\end{equation}

\begin{lemma}
\label{lem:exist-g}
Suppose that~$\tis$ is $n$-maximal
and that $\tir_{i'} \le \tir_{j'}$ whenever $d_{i'} > d_{j'}$.
There exists $w \in [0:\tir_i]$ such that
$\calA_k \ne \emptyset$, where $k = \tir_0 + w \, d_i$.
\end{lemma}

\begin{proof}
Write
$c(x) = \tis(x)/{p_j(x)}^{\tir_j} = a(x) \cdot {p_i(x)}^{\tir_i}$.
We first show that
\begin{equation}
\label{eq:tir0}
0 \le n - \tir_0 \le \deg c .
\end{equation}
Starting with the left inequality in~(\ref{eq:tir0}),
if $\tir_0 > n$ then $\deg s(x) < n$, in which case
\[
\ndiv_n(\tis) = \ndiv(\tis)
< \ndiv(x \cdot \tis(x)) = \ndiv_n(x \cdot \tis(x)) ,
\]
which is impossible since~$\tis$ is $n$-maximal.

Turning to the right inequality in~(\ref{eq:tir0}), observe that
it is equivalent to
\[
\tir_j d_j \le n .
\]
Since $\tir_j > 0$, by our assumptions on $\tis(x)$, this polynomial
has an irreducible factor $p^*(x)$ of degree $d_j - 1$
(taking $p^*(x) = 1$ when $d_j = 1$).
Therefore,
\begin{eqnarray*}
\lefteqn{
\tir_0 + (d_j - 1) + \tir_j d_j + \tir_i d_i
} \makebox[0ex]{} \\
& = & \!\!
\deg \bigl( y^{\tir_0} \cdot p^*(x) \cdot {p_j(x)}^{\tir_j}
\cdot {p_i(x)}^{\tir_i} \bigr)
\le \deg \tis(x,y) = 2 n .
\end{eqnarray*}
But $\tir_0 > 0$, $d_i > d_j$, and
$\tir_i = r_i - 1 \ge r_j = \tir_j - 1$; so,
\[
2 \tir_j d_j \le \tir_0 - 1 + d_j + \tir_j d_j + \tir_i d_i \le 2 n .
\]

Next we turn to constructing
a divisor $g(x)$ of $c(x)$ of degree $n - \tir_0$.
We initialize $\hat{g}(x) \leftarrow 1$.
Then we list the irreducible
factors of $c(x)$ in descending order,
with each factor $p_\ell(x)$
appearing $r_\ell$ times in the list,
and allocate them sequentially to
$\hat{g}(x)$ until one of the following two events occurs
(by~(\ref{eq:tir0}), one of the events must indeed occur):
\begin{itemize}
\item
$\deg \hat{g} = n - \tir_0$.
\item
$\deg \hat{g} < n - \tir_0$, but the next
irreducible factor in the list to be
allocated, $p_\ell(x)$, satisfies
$\deg \hat{g} + d_\ell > n - \tir_0$.
\end{itemize}
In the first case we set $g(x) = \hat{g}(x)$.
In the second case,
we denote $\hat{d} = n - \tir_0 - \deg \hat{g}$
and have $1 \le \hat{d} < d_\ell$.
By our assumptions on~$\tis$, the polynomial $c(x)$ has
a degree-$\hat{d}$ irreducible factor $\hat{p}(x)$
(unless $\hat{d} = d_j = q = 2$, in which case
$p_j(x) = x^2 + x + 1$ is the only irreducible polynomial;
in this case we take $\hat{p}(x) = x(x+1)$,
which divides $c(x)$). From the way $\hat{g}(x)$ is constructed
we have $\gcd(\hat{g},\hat{p}) = 1$, and we define
$g(x) = \hat{g}(x) \cdot \hat{p}(x)$.

Finally, write $g(x) = f(x) \cdot {p_i(x)}^w$,
where $\gcd(f,p_i) = 1$. Then $f(x) \,|\, a(x)$
and $\deg f = n - \tir_0 - w \, d_i = n - k$.
\end{proof}

\begin{lemma}
\label{lem:A(f,g)<=B(f,g)}
For all $k \in [0:\degb]$:
\[
|\calD_k(b)| \le |\calD_k(\tib)| ,
\]
with the inequality being strict when
$k = \tir_0 + w \, d_i$, for any $w \in [0:\tir_i]$.
\end{lemma}

\begin{proof}
Given $k \in [0:\degb]$,
let $\calB$ and $\tilde{\calB}$ be the following subsets
of $\calD_k(b)$ and $\calD_k(\tib)$, respectively:
\begin{eqnarray*}
\calB & = &
\bigl\{ \eta(x,y) \in \calD_k(b) \,:\,
p_i(x) \,|\, \eta(x,y) \bigr\} \\
\tilde{\calB} & = &
\bigl\{ \eta(x,y) \in \calD_k(\tib) \,:\,
y \cdot p_j(x) \,|\, \eta(x,y) \bigr\} .
\end{eqnarray*}
Recalling that $d_j = d_i - 1$
and that $(\tir_0,\tir_j,\tir_i) = (r_0 + 1,r_j + 1,r_i - 1)$, we have:
\begin{eqnarray*}
\lefteqn{
\calB = \bigl\{ y^{w_0} \cdot {p_j(x)}^{w_j}
\cdot {p_i(x)}^{w_i + 1} \,:\,
} \makebox[8ex]{} \\
&&
(w_0,w_j,w_i) \in [0:r_0] \times [0:r_j] \times [0:r_i-1] , \\
&&
\quad
w_0 + w_j (d_i-1)  + w_i d_i = k - d_i
\bigr\}
\end{eqnarray*}
and
\begin{eqnarray*}
\lefteqn{
\tilde{\calB} = \bigl\{ y^{w_0+1} \cdot {p_j(x)}^{w_j+1}
\cdot {p_i(x)}^{w_i} \,:\,
} \makebox[8ex]{} \\
&&
(w_0,w_j,w_i) \in [0:r_0] \times [0:r_j] \times [0:r_i-1] , \\
&&
\quad
w_0 + w_j (d_i-1)  + w_i d_i = k - d_i
\bigr\} ,
\end{eqnarray*}
namely, $|\calB| = |\tilde{\calB}|$.

Next, write
$\calK = \calD_k(b) \setminus \calB$ and
$\tilde{\calK} = \calD_k(\tib) \setminus \tilde{\calB}$;
namely, the elements of $\calK$ may have $y$ and $p_j(x)$
(but not $p_i(x)$) as irreducible factors,
and the elements of $\calK$ cannot have both $y$ and $p_j(x)$
as irreducible factors.
To complete the proof we show that
$|\calK| \le |\tilde{\calK}|$
by verifying that the following mapping
$\varphi: \calK \rightarrow \tilde{\calK}$
is injective:
\[
\varphi \bigl(
y^{w_0} \cdot {p_j(x)}^{w_j} \bigr)
= \left\{
\begin{array}{ll}
{p_j(x)}^{w_j - w_0} \cdot {p_i(x)}^{w_0}
&
\textrm{if $w_0 \le w_j$} \\
y^{w_0 - w_j} \cdot {p_i(x)}^{w_j}
&
\textrm{otherwise} .
\end{array}
\right.
\]
Note that~$\varphi$
is degree-preserving and that
$w_j \le r_j \le r_i - 1 = \tir_i$; so,
$\varphi$ is indeed into $\tilde{\calK}$.
And it is injective with the following inverse:
\begin{eqnarray*}
\varphi^{-1}
\bigl( {p_j(x)}^{\tiw_j} \cdot {p_i(x)}^{\tiw_i} \bigr)
& = &
y^{\tiw_i} \cdot {p_j(x)}^{\tiw_j + \tiw_i} \\
\varphi^{-1}
\bigl( y^{\tiw_0} \cdot {p_i(x)}^{\tiw_i} \bigr)
& = &
y^{\tiw_0 + \tiw_i} \cdot {p_j(x)}^{\tiw_i} .
\end{eqnarray*}
Moreover, for $w \in [0:\tir_i]$
and $k = \tir_0 + w \, d_i$,
the polynomial $y^{\tir_0} \cdot {p_i(x)}^w$ belongs
to $\tilde{\calK}$ yet it is \emph{not} an image of~$\varphi$.
Therefore, $\varphi$ is not surjective
and, so, $|\calK| < |\tilde{\calK}|$.
\end{proof}

\begin{proof}[Proof of \Proposition~\ref{prop:no-zero-holes}]
Suppose that~$s$ is such that $d_i > d_j$ and $r_i > r_j$
for some $i, j \in [1:t]$, and let~$\tis$ be obtained
by~(\ref{eq:bubblesort}).
Combining~(\ref{eq:partitionA}), (\ref{eq:partitionB}),
and Lemma~\ref{lem:A(f,g)<=B(f,g)} yields the (weak)
inequality $\ndiv_n(\tis) \ge \ndiv_n(s)$.
In the remaining part of the proof, we will assume
that~$\tis$ satisfies the condition of the proposition, namely,
that $\tir_i \le \tir_j$ whenever $d_i > d_j$. If it does not,
we can iterate the ``bubble-sort-like'' operation~(\ref{eq:bubblesort})
with~$\tis$ playing the role of~$s$, thereby generating
a sequence of polynomials $s_1 = s, s_2 = \tis_1, s_3 = \tis_2, \ldots$
until the desired condition holds.
Note that the sequence
$(\ndiv_n(s_\ell))_\ell$ is non-decreasing
and that it is finite, since $(\deg s_\ell(x))_\ell$ is decreasing.

If~$\tis$ is not $n$-maximal, then, from
$\ndiv_n(\tis) \ge \ndiv_n(s)$, neither is~$s$.
Otherwise, $\tis$ satisfies the conditions of
Lemma~\ref{lem:exist-g}.
Letting $k = \tir_0 + w \, d_i$ be as in that lemma,
we then have $|\calA_k| > 0$ which,
with (\ref{eq:partitionA}), (\ref{eq:partitionB}),
and Lemma~\ref{lem:A(f,g)<=B(f,g)}, yields the strict inequality
$\ndiv_n(\tis) > \ndiv_n(s)$.
\end{proof}
\fi

\subsection{Proof of \Proposition~\ref{prop:constrained-lowdegrees}}
\label{sec:constrained-lowdegrees}

\ifPAGELIMIT
    To prove \Proposition~\ref{prop:constrained-lowdegrees}
    we first observe that~(\ref{eq:constrained-lowdegrees})
    is equivalent to
    \[
    \rho_n - 2d_i < r_i d_i \le \rho_n.
    \]
    These two inequalities will follow from the next two lemmas.

    \begin{lemma}
    \label{lem:fix-small-ri}
    $r_i d_i > \rho_n-2d_i$ for every $i \in [1:t+1]$.
    \end{lemma}

    \begin{lemma}
    \label{lem:fix-big-ri}
    $r_i d_i < \rho_n + 1$ for every $i \in [1:t]$.
    \end{lemma}

    Fix a polynomial
    $s(x) = \prod_{i=1}^t {p_i(x)}^{r_i} \in \Comp_{2n}$
    and let $s(x,y) = y^{r_0}\cdot s(x)$ be as in~(\ref{eq:s(x,y)}).
    \Withoutlossofgenerality\ assume that $\rho_n = \rho_n(s) = r_0$.

    \begin{proof}[Proof sketch of Lemma~\ref{lem:fix-small-ri}]
    Assuming $r_i d_i \le \rho_n - 2d_i$,
    we show that $\ndiv_n(\tis) > \ndiv_n(s)$, where
    \[
    \tis(x) = s(x) \cdot p_i(x) / y^{d_i}.
    \]
    This is done as follows.
    Write $s(x,y) = a(x) \cdot b(x,y)$
    (as in~(\ref{eq:factorization-a})) and
    $\tis(x) = a(x) \cdot \tib(x,y)$,
    where $b(x,y) = y^{r_0} \cdot {p_i(x)}^{r_i}$
    and $\tib(x,y) = y^{r_0-d_i} \cdot {p_i(x)}^{r_i+1}$
    (and, so, $\gcd(a,b) = \gcd(a,\tib) = 1$),
    $\degb = \deg b(x,y) = r_0 + r_i d_i$,
    and $\deg a(x) = 2n - \degb$.

    First, by combinatorial considerations (which we omit),
    it can be shown that for all $k \in [0:\degb]$:
    \[
    |\calD_k(b)| \le |\calD_k(\tib)| .
    \]
    Then, by a knapsack-like algorithm, one can show
    that $\calA_{\floor{\degb/2}} \ne \emptyset$.
    Finally, one shows that
    \[
    |\calD_{\floor{\degb/2}}(b)| < |\calD_{\floor{\degb/2}}(\tib)|
    \]
    which, with~(\ref{eq:partitionA}) (when stated for~$s$ and~$\tis$)
    yields $\ndiv_n(\tis) > \ndiv_n(s)$.
    \end{proof}

    \begin{proof}[Proof sketch of
    Lemma~\ref{lem:fix-big-ri}]
    Assuming first that $r_i d_i > r_0 + 1$,
    we show that $\ndiv_n(\tis) > \ndiv_n(s)$ for
    \[
    \tis(x) = s(x) \cdot  y^{d_i} / p_i(x) ,
    \]
    similarly to the proof of Lemma~\ref{lem:fix-small-ri}.
    When $r_i d_i = r_0 + 1$, we only get the weak inequality
    $\ndiv_n(\tis) \ge \ndiv_n(s)$,
    yet then $\tis$ (and, therefore, $s$) cannot be $n$-maximal
    since it violates Lemma~\ref{lem:fix-small-ri}.
    \end{proof}
\else
We prove the two inequalities in~(\ref{eq:constrained-lowdegrees})
through a sequence of lemmas.

Fix a polynomial $s(x) = \prod_{i=1}^t {p_i(x)}^{r_i} \in \calP_{2n}$
that satisfies \Proposition~\ref{prop:no-zero-holes}
and let $s(x,y) = y^{r_0}\cdot s(x)$ be as in~(\ref{eq:s(x,y)}).
As was the case in the proof of \Proposition~\ref{prop:no-zero-holes},
we can assume that $\rho_n = \rho_n(s) = r_0$.

Fix also an index $i \in [1:t+1]$.
We will prove that if any of the two inequalities
in~(\ref{eq:constrained-lowdegrees}) does not hold for the selected~$i$,
then~$s$ cannot be $n$-maximal; we do so
(as in previous proofs) by exhibiting
a polynomial~$\tis$ such that $\ndiv_n(\tis) > \ndiv(s)$.

Assume the factorization~(\ref{eq:factorization-a}),
where $b(x,y) =  y^{r_0} \cdot p_i(x)^{r_i}$
(and, therefore, $\gcd(a,b) = 1$),
\[
\degb = \deg b(x,y) = r_0 + r_i d_i ,
\]
and $\deg a(x) = 2n - \degb$.
For every $k \in [0:\degb]$ we define
$\calA_k = \calA_k(n,a)$ as in~(\ref{eq:A}).

\begin{lemma}
\label{lem:exist-f}
$\calA_{\floor{\degb/2}} \ne \emptyset$.
\end{lemma}

\begin{proof}
We construct a divisor $f(x)$ of $a(x)$ of
degree $n-\floor{\degb/2}$
similarly to the construction of $g(x)$ in
the proof of Lemma~\ref{lem:exist-g}.
We initialize $\hat{f}(x) \leftarrow 1$
and then allocate to $\hat{f}(x)$ the irreducible factors of $a(x)$
in descending order until one of the following events occurs:
\begin{itemize}
\item
$\deg \hat{f} = n - \floor{\degb/2}$.
\item
$\deg \hat{f} < n - \floor{\degb/2}$, but the next
irreducible factor to be allocated, $p_\ell(x)$, satisfies
$\deg \hat{f} + d_\ell > n - \floor{\degb/2}$.
\end{itemize}
We proceed as in the proof of Lemma~\ref{lem:exist-g}.
\end{proof}

Turning to the left inequality
in~(\ref{eq:constrained-lowdegrees}),
we assume that it does not hold, namely,
that $r_0 + 1 > (r_i + 2) d_i$,
and---quite similarly to the proof
of \Proposition~\ref{prop:unconstrained-lowdegrees}---we show
that $\ndiv_n(\tis) > \ndiv_n(s)$, where
\[
\tis(x,y) = s(x,y) \cdot p_i(x) / y^{d_i}
= a(x) \cdot \tib(x,y) ,
\]
with
\[
\tib(x,y) = y^{r_0-d_i} \cdot {p_i(x)}^{r_i+1}
\]
(and $\gcd(a,\tib) = 1$)
and $\deg \tib(x,y) = \tir_0 + \tir_i d_i = \degb$;
thus, $\deg \tis(x,y) = \degb + \deg a = \deg s(x,y) = 2n$.

\begin{lemma}
\label{lem:A(f,g)<=B(f,g)-caseA}
If $r_0 + 1 > (r_i + 2) d_i$ then for all $k \in [0:\degb]$:
\[
|\calD_k(b)| \le |\calD_k(\tib)| .
\]
\end{lemma}

\begin{proof}
Since
$|\calD_k(b)| = |\calD_{\degb-k}(b)|$ and
$|\calD_k(\tib)| = |\calD_{\degb-k}(\tib)|$, it
suffices to prove the lemma for $k \le h/2$.
We write $\ell = \degb - k$, where
\[
k \le \floor{\degb/2} \le \ceil{\degb/2} \le \ell.
\]
The size of $\calD_k(b)$ equals the number of ways one can place
$r_i$ identical balls---namely, copies of $p_i(x)$---into
two bins, with at most $\kappa = \floor{k/d_i}$ balls
in the first bin and at most
$\lambda = \floor{\ell/d_i}$ in the second.
One can easily see that
\begin{equation}
\label{eq:|A(f,g)|}
|\calD_k(b)| = \min(r_i,\kappa)- \max(0,r_i - \lambda) + 1 .
\end{equation}
Respectively, with~$b$ and $r_i$ replaced by~$\tib$
and $\tir_i$,
\begin{equation}
\label{eq:|B(f,g)|}
|\calD_k(\tib)|
= \min(\tir_i,\kappa) - \max(0,\tir_i-\lambda)+1.
\end{equation}
Now, the assumption $r_0 + 1 > (r_i + 2) d_i$ implies
\begin{eqnarray}
&&
\tir_i d_i = (r_i + 1)d_i \le r_0 - d_i
=\tir_0 \nonumber\\
\label{ineq:ridi+di<=m_max}
& \Rightarrow &
\tir_i d_i \le
\floor{(\tir_0 + \tir_i d_i)/2}
= \floor{\degb/2} \le \ell \\
& \Rightarrow &
\tir_i \le
\floor{\ell/d_i} = \lambda \nonumber\\	
\label{eq:second-part-in-|A(f,g)|-caseA}
& \Rightarrow &
\max(0,r_i - \lambda) = \max(0,\tir_i-\lambda)= 0.
\end{eqnarray}
On the other hand, $r_i < \tir_i$ implies
$\min(r_i,\kappa) \le \min(\tir_i,\kappa)$.
Combining this with (\ref{eq:|A(f,g)|})--(\ref{eq:|B(f,g)|})
and~(\ref{eq:second-part-in-|A(f,g)|-caseA}) leads to
\[
|\calD_k(b)| \le |\calD_k(\tib)| .
\]\end{proof}

\begin{proof}[Proof of
               the left inequality in~(\ref{eq:constrained-lowdegrees})]
We show that if $r_0 + 1 > (r_i + 2) d_i$
then $\ndiv_n(\tis) > \ndiv_n(s)$
(and, so, $s$ cannot be $n$-maximal).
Combining~(\ref{eq:partitionA})
(when stated for~$s$ and~$\tis$)
with Lemma~\ref{lem:A(f,g)<=B(f,g)-caseA}
yields the weak inequality
$\ndiv_n(\tis) \ge \ndiv_n(s)$.
To obtain the strict inequality, we consider
the case $k = \floor{\degb/2}$:
by Lemma~\ref{lem:exist-f} we have $|\calA_k| > 0$,
and we will show that
\[
|\calD_k(b)| < |\calD_k(\tib)| .
\]
As we saw in~(\ref{ineq:ridi+di<=m_max}),
the assumption $r_0 + 1 > (r_i + 2) d_i$ implies
\begin{eqnarray*}
&&
\tir_i d_i \le \floor{(r_0 + r_id_i)/2}
= \floor{\degb/2} = k \\
& \Rightarrow &
r_i < \tir_i \le \floor{k/d_i} = \kappa,
\end{eqnarray*}
hence $\min(r_i,\kappa) = r_i$ and
$\min(\tir_i,\kappa) = \tir_i$.
Combining this with~(\ref{eq:|A(f,g)|})--(\ref{eq:|B(f,g)|})
and~(\ref{eq:second-part-in-|A(f,g)|-caseA})
leads to
$|\calD_k(b)| = r_i + 1 < \tir_i + 1 = |\calD_k(\tib)|$.
\end{proof}

Turning next to the right inequality
in~(\ref{eq:constrained-lowdegrees}),
we again assume that it does not hold, namely,
that $r_0 + 1 \le r_i d_i$, and define
\[
\tis(x,y) = s(x,y) \cdot y^{d_i} / p_i(x)
= a(x) \cdot \tib(x,y) ,
\]
where
\[
\tib(x,y) = y^{r_0 + d_i} \cdot {p_i(x)}^{r_i-1} .
\]
Here, too,
$\deg \tib(x,y) = \tir_0 + \tir_i d_i = \degb$
and, so, $\deg \tis(x,y) = \deg s(x,y) = 2n$.

\begin{lemma}
\label{lem:A(f,g)<=B(f,g)-caseB}
If $r_0 + 1 \le r_i d_i$ then for all $k \in [0:\degb]$:
\[
|\calD_k(b)| \le |\calD_k(\tib)| .
\]
\end{lemma}

\begin{proof}
Using the notation
$\kappa = \floor{k/d_i}$ and $\lambda = \floor{\ell/d_i}$
as in the proof of Lemma~\ref{lem:A(f,g)<=B(f,g)-caseA},
we note that~(\ref{eq:|A(f,g)|}) and~(\ref{eq:|B(f,g)|}) still hold.
The assumption $r_0 + 1 \le r_i d_i$ then implies
\begin{eqnarray}
&&
k \le \floor{\degb/2} = \floor{(r_0 + r_i d_i)/2} < r_i d_i \nonumber\\
& \Rightarrow &
\kappa = \floor{k/d_i}
\le \tir_i < r_i \nonumber\\
& \Rightarrow &
\min(r_i,\kappa)=\min(\tir_i,\kappa) = \kappa.
\label{eq:first-part-in-|A(f,g)|-caseB}
\end{eqnarray}
On the other hand, $\tir_i < r_i$ implies
$\max(0,\tir_i - \lambda) \le \max(0,r_i-\lambda)$.
Combining this with~(\ref{eq:|A(f,g)|})--(\ref{eq:|B(f,g)|})
and~(\ref{eq:first-part-in-|A(f,g)|-caseB}) leads to
$|\calD_k(b)| \le |\calD_k(\tib)|$.
\end{proof}

\begin{proof}[Proof of
              the right inequality in~(\ref{eq:constrained-lowdegrees})]
Assuming that $r_0 + 1 \le r_i d_i$,
we show that~$s$ cannot be $n$-maximal;
note that Eq.~(\ref{eq:partitionA}) (when stated for~$s$ and~$\tis$)
and Lemma~\ref{lem:A(f,g)<=B(f,g)-caseB} already yield
the weak inequality $\ndiv_n(\tis) \ge \ndiv_n(s)$.
We distinguish between two cases.

\emph{Case~1: $r_0 + 1 < r_i d_i$.}
Letting $k = \floor{\degb/2}$, we show that
\[
|\calD_k(b)| < |\calD_k(\tib)| ,
\]
thereby leading, along with Lemma~\ref{lem:exist-f},
to the strong inequality $\ndiv_n(\tis) > \ndiv_n(s)$.
The assumption $r_0 + 1 < r_i d_i$ implies
\[
\ell = \ceil{\degb/2} = \ceil{(r_0 + r_i d_i)/2} < r_i d_i .
\]
Thus, $\lambda = \floor{\ell/d_i} \le \tir_i < r_i$, so we get that
\begin{equation}
\label{eq:second-part-in-|A(f,g)|-caseB}
\max(0,r_i - \lambda) = r_i - \lambda, \qquad
\max(0,\tir_i - \lambda) = \tir_i-\lambda.
\end{equation}
Therefore, (\ref{eq:|A(f,g)|})--(\ref{eq:|B(f,g)|})
and
(\ref{eq:first-part-in-|A(f,g)|-caseB})--%
                                (\ref{eq:second-part-in-|A(f,g)|-caseB})
can be combined to obtain
\[
|\calD_k(b)| = \kappa - r_i + \lambda + 1
< \kappa - \tir_i + \lambda + 1 = |\calD_k(\tib)| .
\]

\emph{Case~2: $r_0 + 1 = r_i d_i$.}
We proceed similarly to the proof of
\Proposition~\ref{prop:unconstrained-lowdegrees}.
In this case $d_i \ge 2$ and, so,
for any $p_j(x) \in \calM_1$:
\[
\tilde{\rho}_n = \rho(\tis,n) = r_0 + d_i \ge r_j + 2 = \tir_j + 2 ,
\]
namely,
\[
\tilde{\rho}_n + 1 > \tir_j + 2 = (\tir_j + 2) d_j ,
\]
which means that~$\tis$ does not satisfy
the left inequality in~(\ref{eq:constrained-lowdegrees})
and therefore is not $n$-maximal.
Yet $\ndiv_n(\tis) \ge \ndiv_n(s)$,
so~$s$ is not $n$-maximal either.
\end{proof}
\fi

The counterparts of Lemma~\ref{lem:rmax-d_t-crude}
and \Proposition~\ref{prop:rmax-d_t-crude}
for $n$-maximal
\ifPAGELIMIT
    polynomials, which are obtained by replacing~$m$ and~$\rho$ therein
    by $2n$ and $\rho_n$, respectively,
\else
polynomials take the form
\begin{equation}
\label{ineq:r-lower-upper-bounds-by-dt}
d_t \le \rho_n \le 2d_{t+1} - 1
\end{equation}
and
\begin{equation}
\label{eq:d_t=log_q(n)}
\floor{\log_q (n/4)} < d_t \le d_{t+1} \le \floor{\log_q (2n)} + 1 ,
\end{equation}
and
\fi
are proved similarly.

\subsection{Proof of
\Proposition~\ref{prop:unconstrained-rmaxexact}
and \Theorem~\ref{thm:unconstrained-highdegrees}
for the $n$-maximal case}
\label{sec:unconstrained-highdegrees}

In this section, we show
that \Proposition~\ref{prop:unconstrained-rmaxexact}
and \Theorem~\ref{thm:unconstrained-highdegrees}
hold also for the $n$-maximal case.

Fix an $n$-maximal polynomial
$s(x) = \prod_{i=1}^t {p_i(x)}^{r_i}$,
let $s(x,y) = y^{r_0}\cdot s(x)$
where $r_0 = 2n - \deg s(x)$, and write $\rho_n = \rho_n(s)$.
\ifPAGELIMIT
    Fix a factorization~(\ref{eq:factorization-a})
    where $\gcd(a,b) = 1$ and $b(x,y) \in \Comp_\degb$,
    for some $\degb \in [r_0:2n]$.
    For every $k \in [\degb{-}n:n]$ let
    $\calA_k = \calA_k(n,a)$ be as in~(\ref{eq:A}).
\else
We assume hereafter that $n \ge 4 q^2$ which,
by (\ref{ineq:r-lower-upper-bounds-by-dt})--(\ref{eq:d_t=log_q(n)}),
implies that $d_{t+1}, \rho_n \ge d_t > 2$.

Fix a factorization~(\ref{eq:factorization-a})
where $\gcd(a,b) = 1$ and $b(x,y) \in \Comp_\degb$,
for some $\degb \in [r_0:2n]$.
For every $k \in [\degb{-}n:n]$ let
$\calA_k = \calA_k(n,a)$ be as in~(\ref{eq:A})
and $\calA = \bigcupdot_{k \in [\degb{-}n:n]} \calA_k$
be the set of divisors of $a(x)$.

The following proposition specifies a range of values of~$\degb$
(that will suffice for our purposes)
for which the size of $\calA_k$ varies very little over
$k \in [0:\degb]$.
(In all the $\O (\cdot)$ terms hereafter, the multiplying constants
are absolute, namely, independent of~$q$ and~$n$.)
\fi

\begin{proposition}
\label{prop:flat}
Let $s(x,y) \in \Comp_{2n}$ be $n$-maximal and assume
the factorization~(\ref{eq:factorization-a})
with $\degb = \deg b(x,y) = \O(\log_q^2 n)$.
For any $k, k' \in [0:\degb]$:
\ifPAGELIMIT
    \[
    |\calA_{k'}|/|\calA_k|
    \ge 1 - \O \left( \lambda_q(n) \right) ,
    \]
    where $\lambda_q(n) = \sqrt{(q \ln n)/n} \cdot \log_q^2 n$.
\else
\begin{equation}
\label{eq:flat}
\frac{|\calA_{k'}|}{|\calA_k|} \ge 1 - \O \left( \lambda_q(n)\right) ,
\end{equation}
where
\[
\lambda_q(n) = \sqrt{\frac{q \ln n}{n}} \cdot \log_q^2 n.
\]
\fi
\end{proposition}

\ifPAGELIMIT
    We omit the proof of the proposition due to space limitations.
    Next, we sketch how it
    implies Lemma~\ref{lem:diupper}(a)
    (and, similarly, Lemma~\ref{lem:diupper}(b)
    and Lemma~\ref{lem:dilower}(a)--(b)
    and, consequently,
    \Proposition~\ref{prop:unconstrained-rmaxexact}
    and \Theorem~\ref{thm:unconstrained-highdegrees})
    for the $n$-maximal case.
\else
We prove the proposition in Section~\ref{sec:flat} below.
Before doing so, we demonstrate how it implies
\Proposition~\ref{prop:unconstrained-rmaxexact}
and \Theorem~\ref{thm:unconstrained-highdegrees}
for the $n$-maximal case, by inserting
slight changes into the proofs of
Lemmas~\ref{lem:diupper} and~\ref{lem:dilower}
(we will show the change for Lemma~\ref{lem:diupper}(a)
and its effect on \Proposition~\ref{prop:unconstrained-rmaxexact};
the other changes are similar).
\fi
Assuming that $s(x)$ is $n$-maximal, we define
the sets~$\Set$ and~$\Setalt$ 
and the polynomial $\tis(x)$ as in the proof
of Lemma~\ref{lem:diupper}(a).
We write $s(x,y) = y^{r_0} \cdot s(x) = a(x) \cdot b(x,y)$, where
\[
\ifPAGELIMIT
    \textstyle
\fi
b(x,y) = y^{r_0} \cdot \prod_{j \in \Set} {p_j(x)}^{r_j} .
\]
Similarly, we write
$\tis(x,y) = y^{r_0} \cdot \tis(x) = a(x) \cdot \tib(x,y)$,
where
\[
\ifPAGELIMIT
    \textstyle
\fi
\tib(x,y) = y^{r_0} \cdot \prod_{k \in \Setalt} p_k(x)
\cdot \prod_{j \in \Set} {p_j(x)}^{r_j-1} .
\]
The degree $\degb = \deg b(x,y) = \deg \tib(x,y)$ is given by
\[
\ifPAGELIMIT
    \textstyle
\fi
\degb = r_0 + \sum_{j \in \Set} r_j d_j
\le r_0 + \rho_n (d_t + 1) = \O \left( \log_q^2 n \right) .
\]
Denoting
\[
\calD(b) = \bigcupdot_{k \in [0:\degb]} \calD_k(b)
\ifIEEE
    \quad \textrm{and} \quad
\else
\quad\quad \textrm{and} \quad\quad
\fi
\calD(\tib) = \bigcupdot_{k \in [0:\degb]} \calD_k(\tib)
\ifPAGELIMIT
    ,
\fi
\]
\ifPAGELIMIT
\else
(the sets of divisors of~$b$ and~$\tib$, respectively),
\fi
we recall that, by~(\ref{eq:factorization}),
\begin{eqnarray}
\label{eq:calR}
|\calD(b)|
& = &
\ifPAGELIMIT
    \textstyle
\fi
(r_0 + 1) \cdot \prod_{j \in \Set} (r_j + 1) \\
\label{eq:tildecalR}
|\calD(\tib)|
& = &
\ifPAGELIMIT
    \textstyle
\fi
(r_0 + 1) \cdot 2^{d_i} \cdot \prod_{j \in \Set} r_j .
\end{eqnarray}
  From~(\ref{eq:partitionA}) (when stated for~$s$ and~$\tis$),
(\ref{eq:calR})--(\ref{eq:tildecalR}),
and \Proposition~\ref{prop:flat} we get:
\begin{eqnarray*}
\frac{\ndiv_n(\tis)}{\ndiv_n(s)}
& \ge &
\frac{\min_{k \in [0:\degb]} |\calA_k|}%
                        {\max_{k \in [0:\degb]} |\calA_k|}
\cdot
\frac{\sum_{k \in [0:\degb]} |\calD_k(\tib)|}%
                        {\sum_{k\in [0:\degb]} |\calD_k(b)|} \\
& \ge &
\left( 1 - \O \left( \lambda_q(n) \right) \right)
\cdot \frac{|\calD(\tib)|}{|\calD(b)|} \\
\ifPAGELIMIT
\else
& = &
\left( 1 - \O \left( \lambda_q(n) \right) \right)
\cdot
2^{d_i} \cdot
\prod_{j \in \Set}
\left( \frac{r_j}{r_j + 1} \right) \\
\fi
& \ge &
\left( 1 - \O \left( \lambda_q(n) \right) \right)
\cdot
2^{d_i} \cdot
\left( \frac{r_i - 1}{r_i} \right)^{d_t + 1} ,
\end{eqnarray*}
which is the same as~(\ref{eq:diupper2}) except for
the multiplicative $1 - \O \left( \log_q^2 n \right)$ term.
Taking logarithms, we will have
an $\O \left( \lambda_q(n) \right)$ term
subtracted from the left-hand side of~(\ref{eq:diupper1})
and, consequently, from each instance of $d_i$
in~(\ref{eq:rmaxexact1}).
Since this term goes to zero as $n \rightarrow \infty$
much faster than $d_i/d_t$, its contribution amounts to
adding an $o_n(1)$ term to the upper bound~(\ref{eq:rmaxexact2}).

\ifPAGELIMIT
\else
\subsection{Proof of \Proposition~\ref{prop:flat}}
\label{sec:flat}

We prove \Proposition~\ref{prop:flat} through a sequence of
definitions and lemmas.

For $d \in [1:d_t]$, write $r_+(d) = \floor{\rho_n/d}$
and $r_-(d) = r_+(d) - 1$, and define $\N_+(d)$ and $\N_-(d)$ by
\[
\N_\pm(d) =
\left|
\bigl\{
p_i(x) \,:\, p_i \,|\, a , \,
\deg p_i = d , \, \mult_{p_i}(a) = r_\pm(d)
\bigr\}
\right| ,
\]
namely, $\N_\pm(d)$ is the number of
distinct irreducible factors of $a(x)$ of degree~$d$
and of multiplicity $r_\pm(d)$
(by \Proposition~\ref{prop:constrained-lowdegrees},
$r_\pm(d)$ are the only possible multiplicities of such factors).
We have
\begin{equation}
\label{eq:degree-a}
\sum_{(d,\sigma) \in [1:d_t] \times \{ \pm \}}
d \cdot r_\sigma(d) \cdot \N_\sigma(d) = \deg a(x) = 2n - \degb
\end{equation}
and
\begin{equation}
\label{eq:sumNs}
\N_+(d) + \N_-(d) \le \I(d) ,
\end{equation}
with equality holding
(by \Proposition~\ref{prop:no-zero-holes}) for all~$d$,
except when $b(x,y)$
has irreducible factors of degree~$d$ or when
$d = d_t = d_{t+1}$.

A \emph{type} is a list $\bldtheta$ of nonnegative integers of the form
\begin{equation}
\label{eq:type1}
\bldtheta =
\bigl(
\N_\sigma(d,0), \N_\sigma(d,1), \ldots, \N_\sigma(d,r_\sigma(d))
\bigr)_{(d,\sigma) \in [1:d_t] \times \{ \pm \}} ,
\end{equation}
where for each $(d,\sigma) \in [1:d_t] \times \{ \pm \}$:
\begin{equation}
\label{eq:type2}
\sum_{r=0}^{r_\sigma(d)} \N_\sigma(d,r) = \N_\sigma(d) .
\end{equation}
Denoting by $L = L(a)$ the number of different types,
we have the following lemma.

\begin{lemma}
\label{lem:typeenumeration}
\[
L \le n^{14.5 + 3.5 \log_q n} .
\]
\end{lemma}

\begin{proof}
It is easy to see that
\[
L
\le
\prod_{(d,\sigma) \in [1:d_t] \times \{ \pm \}}
{\left( \N_\sigma(d) + 1 \right)}^{r_\sigma(d)} .
\]
By the AM--GM inequality we have, for every $d \in [1:d_t]$:
\begin{eqnarray*}
\prod_{\sigma \in \{ \pm \}}
{\left( \N_\sigma(d) + 1 \right)}^{r_\sigma(d)}
\!
& \stackrel{(\ref{eq:sumNs})}{\le} &
{\left( \frac{\I(d){+}1}{2} \right)}^{2 \, r_-(d)}
\cdot
(\I(d){+}1) \\
& \le &
\frac{1}{2^{2 \, r_-(d)}} \cdot
{\left( \frac{q^d}{d} + 1 \right)}^{2 \, r_-(d) + 1} \\
& \le &
q^{d \cdot (2 \, r_-(d) + 1)} ,
\end{eqnarray*}
where the last inequality holds whenever $d > 1$ or
$r_-(d) > 0$, and,
by \Proposition~\ref{prop:constrained-lowdegrees},
we indeed have $r_-(1) > 0$ since $\rho_n > 2$.
Hence,
\begin{eqnarray*}
L
& \le &
\prod_{d \in [1:d_t]}
q^{2 d \, r_+(d) - d}
\le
\prod_{d \in [1:d_t]}
q^{2 \rho_n - d} \\
& \le &
q^{(2 \rho_n - (d_t+1)/2) d_t}
\le
{\left( 2 q n \right)}^{3.5 \log_q n + 5},
\end{eqnarray*}
where the last step follows from
(\ref{ineq:r-lower-upper-bounds-by-dt})--(\ref{eq:d_t=log_q(n)})
and $d_t \ge d_{t+1} - 1$. Recalling our assumption that
$q \le \sqrt{n}/2$, we finally get:
\begin{eqnarray*}
L
\le
{\left( 2 q n \right)}^{3.5 \log_q n + 5}
& \le &
{\left( q^2 n \right)}^{3.5 \log_q n}
\cdot {\left( 2 q n \right)}^5 \\
& \le &
n^{14.5 + 3.5 \log_q n} .
\end{eqnarray*}
\end{proof}

Given a divisor $f \in \calA$ of $a(x)$,
we denote by $\Type{f}$ the type~$\bldtheta$
as in~(\ref{eq:type1})--(\ref{eq:type2}), where
\begin{eqnarray*}
\lefteqn{
\N_\pm(d,r) =
\bigl|
\bigl\{
p_i(x) \,:\, p_i \,|\, f , \;
\deg p_i = d ,
} \makebox[10ex]{} \\
&&
\quad\quad
\mult_{p_i}(f) = r , \;
\mult_{p_i}(a) = r_\pm(d)
\bigr\}
\bigr| ,
\end{eqnarray*}
namely, $\N_\sigma(d,r)$ is the number of degree-$d$
irreducible factors of $f(x)$ that have multiplicities~$r$
and $r_\sigma(d)$ in $f(x)$ and $a(x)$, respectively.

For any type $\bldtheta$ as in~(\ref{eq:type1})--(\ref{eq:type2}),
we define
\[
\calA(\bldtheta) =
\left\{ f \in \calA \,:\, \Type{f} = \bldtheta \right\} .
\]
We can generate any element $f \in \calA(\bldtheta)$
by selecting its irreducible factors and their respective
multiplicities as follows.
For each $(d,\sigma) \in [1:d_t] \times \{ \pm \}$,
partition the $N_\sigma(d)$ degree-$d$ irreducible factors
of multiplicity $r_\sigma(d)$ of $a$ into $r_\sigma+1$ bins
so that each bin $r \in [0:r_\sigma]$ contains
$N_\sigma(d,r)$ factors; the factors in bin $r \in [1:r_\sigma]$
are then taken to be irreducible factors of multiplicity $r$ in $f$.
Doing so, we see that
the size of $\calA(\bldtheta)$ is given by the following
product of multinomial coefficients:
\begin{equation}
\label{eq:Ltau}
\left| \calA(\bldtheta) \right|
=
\prod_{(d,\sigma) \in [1:d_t] \times \{ \pm \}}
\frac{\N_\sigma(d)!}{\prod_{r \in [0:r_\sigma(d)]} \N_\sigma(d,r)!} .
\end{equation}
The \emph{degree} of~$\bldtheta$, denoted $\deg \bldtheta$,
is the degree of each $f \in \calA(\bldtheta)$:
\begin{equation}
\label{eq:degreetau}
\deg \bldtheta =
\sum_{(d,\sigma) \in [1:d_t] \times \{ \pm \}}
d \cdot
\sum_{r \in [0:r_\sigma(d)]}
r \cdot \N_\sigma(d,r) .
\end{equation}

The next two lemmas characterize types $\bldtheta$ for which
$\left| \calA(\bldtheta) \right|$ is maximized.

\begin{lemma}
\label{lem:maximaltype}
The size of $\calA(\bldtheta)$ is maximized for any type~$\bldtheta$
that satisfies:
\[
\left|
\N_\sigma(d,r) - \frac{\N_\sigma(d)}{r_\sigma(d) + 1} \right| < 1
\]
for every $(d,\sigma) \in [1:d_t] \times \{ \pm \}$
and $r \in [0:r_\sigma(d)]$.
\end{lemma}

\begin{proof}
By the known properties of the multinomial coefficients,
for each pair $(d,\sigma)$, the respective term in~(\ref{eq:Ltau})
is maximized when (and only when)
$\N_\sigma(d,r)$ is either the floor or the ceiling
of $\N_\sigma(d)/(r_\sigma(d) + 1)$,
subject to the constraint~(\ref{eq:type2}).
\end{proof}

\begin{lemma}
\label{lem:maximalk}
Among the maximizing types in Lemma~\ref{lem:maximaltype},
there exists a type~$\bldtheta_0$ such that
\[
\Bigl| \deg \bldtheta_0 - n + \frac{\degb}{2} \Bigr|
\le \frac{\rho_n}{2} .
\]
\end{lemma}

\begin{proof}
Let $\bldtheta$ in~(\ref{eq:type1})--(\ref{eq:type2}) be
an (initial) maximizing type,
and for some $(\bar{d},\bar{\sigma}) \in [1:d_t] \times \{ \pm \}$
and $\bar{r} \in [0:\floor{r_{\bar{\sigma}}(\bar{d})/2}]$,
define the type $\bar{\bldtheta}$ by ``switching''
a pair of values in $\bldtheta$ as follows:
\[
\bar{\N}_\sigma(d,r) =
\left\{
\begin{array}{ll}
\N_\sigma(d,r_{\sigma}(d)-r)
& \textrm{if $(d,\sigma) = (\bar{d},\bar{\sigma})$ and} \\
& \;\;\quad \textrm{$r \in \{ \bar{r}, r_{\sigma}(d) - \bar{r} \}$} \\
\N_\sigma(d,r)
& \textrm{otherwise} .
\end{array}
\right.
\]
The type $\bar{\bldtheta}$ is also maximizing and
\begin{eqnarray*}
\ifIEEE
    \bigl| \deg \bar{\bldtheta} - \deg \bldtheta \bigr|
    & = & \bar{d} \cdot
    \bigl| \left( r_{\bar{\sigma}}(\bar{d}) - 2 \bar{r} \right) \\
    &&
    \quad {}
    \cdot
    \left(\N_{\bar{\sigma}}(\bar{d},r_{\bar{\sigma}}(\bar{d})-\bar{r})
    - \N_{\bar{\sigma}}(\bar{d},\bar{r}) \right) \bigr| \\
\else
\bigl| \deg \bar{\bldtheta} - \deg \bldtheta \bigr|
& = & \bar{d} \cdot
\bigl| \left( r_{\bar{\sigma}}(\bar{d}) - 2 \bar{r} \right)
\cdot
\left(\N_{\bar{\sigma}}(\bar{d},r_{\bar{\sigma}}(\bar{d})-\bar{r})
- \N_{\bar{\sigma}}(\bar{d},\bar{r}) \right) \bigr| \\
\fi
& \le &
\bar{d} \cdot r_{\bar{\sigma}}(\bar{d}) \le \rho_n .
\end{eqnarray*}
If we now start with~$\bldtheta$ and perform \emph{all} such possible
switches one by one, we will end up with a maximizing
type~$\bldtheta'$ with degree
\[
\deg \bldtheta' =
\sum_{(d,\sigma) \in [1:d_t] \times \{ \pm \}}
d \cdot
\sum_{r \in [0:r_\sigma(d)]} (r_\sigma(d) - r) \cdot \N_\sigma(d,r) ,
\]
and, so, by~(\ref{eq:degree-a}), (\ref{eq:type2}),
and~(\ref{eq:degreetau})
we have $\deg \bldtheta + \deg \bldtheta' = 2n - \degb$.
We conclude that either
$\deg \bldtheta \le n - (\degb/2) \le \deg \bldtheta'$
or both inequalities are reversed.
Hence, as we iterate over the switches,
the sequence of degrees of the generated types,
which change at each step by at most $\rho_n$, must at some point cross
the value $n - (\degb/2)$. The type just before or just after
this crossing point is the desired type $\bldtheta_0$.
\end{proof}

Hereafter, we fix $\bldtheta_0$ to be a maximizing type
as in Lemma~\ref{lem:maximalk}.

For $d \in [1:d_t]$, we denote by $\sigma(d)$
a value $\sigma \in \{ \pm \}$ for which
$\N_\sigma(d) \ge \N_{-\sigma}(d)$.
We will use the short-hand notation
$\N(d) = \N_{\sigma(d)}(d)$ and
$r(d) = r_{\sigma(d)}(d)$, and extend this convention also
to any type $\bldtheta$ in writing $\N(d,r) = \N_{\sigma(d)}(d,r)$.
Also, define $\delta$ as follows:
\[
\delta =
\left\{
\begin{array}{ll}
d_t
& \textrm{if $\N(d_t) \ge \N(d_t - 1)$} \\
d_{t+1} - 1
& \textrm{otherwise} .
\end{array}
\right.
\]

\begin{lemma}
\label{lem:Ndelta}
Assuming that $\degb = o \left( n/(q \log_q n) \right)$,
\[
\N(\delta) = \Theta \left( n/\log_q n \right)
\ifIEEE
    \quad \textrm{and} \quad
\else
\quad\quad \textrm{and} \quad\quad
\fi
\N(\delta - 1) = \Omega \left( n/(q \log_q n) \right) .
\]
\end{lemma}

\begin{proof}
Following similar arguments as in
the proof of \Theorem~\ref{thm:upper-bound},
the number, $w_1$, of the irreducible factors of~$s$
of degree at most $\Delta = \floor{(1/2) \log_q n}$
(counting multiplicities) is $\O \left( n^{1/2} \right)$.
The number, $w_2$, of the remaining irreducible factors
is at least $(2n-w_1 \Delta)/d_t$ and at most $2n/\Delta$, namely,
$w_2 = \Theta \left( n/\log_q n \right)$;
moreover, by \Proposition~\ref{prop:constrained-lowdegrees}
and Eq.~(\ref{ineq:r-lower-upper-bounds-by-dt}),
the multiplicity of each of these factors is at most
$\rho_n/\Delta = \O \left( 1 \right)$.
We also recall from \Proposition~\ref{prop:no-zero-holes}
that for $d \le d_{t+1} - 1$,
the number of distinct irreducible factors of $s(x)$
of degree~$d$ is $\I(d) = \Theta \left( q^d/d \right)$.
Hence,
\begin{equation}
\label{eq:Ndelta1}
\N(\delta) \ge \N(\delta - 1)
= \Omega \left( \N(\delta)/q - \degb \right)
\end{equation}
and (by~(\ref{eq:J(d)-estimation}))
\begin{equation}
\label{eq:Ndelta2}
\N(\delta) + \N(\delta - 1) = \Theta \left( w_2 - \degb \right)
= \Theta \left( n/\log_q n \right) .
\end{equation}
The result follows from~(\ref{eq:Ndelta1}) and~(\ref{eq:Ndelta2}).
\end{proof}

A type $\bldtheta$ is called \emph{balanced} if
for each $d \in \{ \delta, \delta - 1 \}$
and $r \in \{ 0, 1 \}$:
\begin{equation}
\label{eq:balanced}
\left|
\frac{\N(d,r)}{\N(d)} - \frac{1}{r(d) + 1} \right|
\le \frac{\gamma_q(n)}{\sqrt{\N(d)}} ,
\end{equation}
where
\[
\gamma_q(n) = \sqrt{6 \ln (n) \cdot \log_q n} .
\]
Note that for $d \in \{ \delta, \delta - 1 \}$
(and $d_{t+1} > 2$) we have
\[
r(d) \le r_+(d)
= \floor{\frac{\rho_n}{d}}
\le \floor{\frac{2d_{t+1} - 1}{d_{t+1} - 2}} \le 5 .
\]

\begin{lemma}
\label{lem:not-balanced}
If $\bldtheta$ is \underline{not} balanced, then
\[
\left| \calA(\bldtheta) \right|
= \O \left( n^{2.5 - 12 \log_q n} \right) \cdot
\left| \calA(\bldtheta_0) \right| .
\]
\end{lemma}

\begin{proof}
Suppose that~(\ref{eq:balanced})
does not hold for some
$(d,r') \in \{ \delta, \delta - 1 \} \times \{ 0, 1 \}$.
Let
$R_1, R_2, \ldots, R_{\N(d)}$ be i.i.d.\ random variables
with $\Prob \left\{ R_j = r \right\} = \pi = 1/(r(d) + 1)$ for each
$j \in [1:\N(d)]$ and $r \in [0:r(d)]$.
Denoting
\[
S(r) = \left| \left\{ j \,:\, R_j = r \right\} \right| ,
\]
we have:
\begin{eqnarray*}
\ifIEEE
    \lefteqn{
    \frac{\N(d)!}{\prod_{r \in [0:r(d)]} \N(d,r)!}
    \cdot \pi^{\N(d)}
    } \makebox[10ex]{} \\
\else
\frac{\N(d)!}{\prod_{r \in [0:r(d)]} \N(d,r)!}
\cdot \pi^{\N(d)}
\fi
& = &
\Prob \left\{
{\textstyle\bigcap_{r \in [0:r(d)]}} \bigl( S(r) = \N(d,r) \bigr)
\right\} \\
& \le &
\Prob \left\{
\left| \frac{S(r')}{\N(d)} - \pi \right|
> \frac{\gamma_q(n)}{\sqrt{\N(d)}} \right\} \\
& \le &
2 \, e^{-2 \, {\gamma_q(n)}^2} ,
\end{eqnarray*}
where the last step follows from
Hoeffding's inequality~\cite[\Theorem~1]{Hoeffding}.
Hence,
\begin{eqnarray*}
\frac{\N(d)!}{\prod_{r \in [0:r(d)]} \N(d,r)!}
& \le &
2 \, e^{-2 \, {\gamma_q(n)}^2} \cdot {(r(d) + 1)}^{\N(d)} \\
& = &
2 \, n^{-12 \log_q n} \cdot {(r(d) + 1)}^{\N(d)} .
\end{eqnarray*}
On the other hand, the respective term in
the expression~(\ref{eq:Ltau})
for $\left| \calA(\bldtheta_0) \right|$ equals
\[
\frac{\N(d)!}{\prod_{r \in [0:r(d)]} \N(d,r)!}
= \Theta \left( {\N(d)}^{-r(d)/2} \right)
\cdot {(r(d) + 1)}^{\N(d)} ,
\]
where we have used the Stirling approximation for the binomial
coefficients (see, for example~\cite[p.~309, Eq.~(16)]{MS}).
The result follows by recalling that
$r(d) \le 5$
and (from the proof of Lemma~\ref{lem:Ndelta})
that $\N(d) = \O \left( n/\log_q n \right)$.
\end{proof}

Given an integer $k \in [\degb{-}n:n]$, we say that the set
$\calA_k$ (as in~(\ref{eq:A})) is \emph{rich} if
$|\calA_k| \ge (1/2) \left| \calA(\bldtheta_0) \right|$.

\begin{lemma}
\label{lem:balanced-k}
Assuming that $\degb = o \left( n/(q \log_q n) \right)$,
let $k \in [d_t{+}\degb{-}n:n{-}d_t]$ be such that $\calA_k$ is rich.
Then for $d \in \{ \delta, \delta - 1 \}$,
\[
\frac{|\calA_{k \pm d}|}{|\calA_k|}
\ge 1 - \O \left( \lambda_q(n) / \log_q n \right) .
\]
\end{lemma}

\begin{proof}
We prove the lemma when stated with the plus sign;
the other case is similar.
Let $\calT$ denote the set of all balanced types~$\bldtheta$
such that $\calA(\bldtheta) \subseteq \calA_k$.
By Lemmas~\ref{lem:typeenumeration} and~\ref{lem:not-balanced} we have
\begin{eqnarray}
\sum_{\bldtheta \in \calT} \left| \calA(\bldtheta) \right|
& \ge &
|\calA_k|
- \O \left( n^{2.5 - 12 \log_q n} \right) \cdot
L \cdot \left| \calA(\bldtheta_0) \right|
\nonumber \\
\label{eq:balanced-k1}
& \ge &
\left( 1 - \O \left( n^{17 - 8.5 \log_q n} \right) \right)
\cdot |\calA_k| .
\end{eqnarray}
Next, for each type $\bldtheta \in \calT$, we associate,
in a one-to-one manner, a type $\varphi(\bldtheta)$ obtained by
adding~$1$ to $\N(d,0)$ and subtracting~$1$ from $\N(d,1)$.
It is easy to see that
$\calA(\varphi(\bldtheta)) \subseteq \calA_{k+d}$ and that
\begin{eqnarray}
\nonumber
\frac{\left| \calA(\varphi(\bldtheta))
\right|}{\left| \calA(\bldtheta) \right|}
& = &
\frac{\N(d,1)}{\N(d,0) + 1} \\
\nonumber
& \stackrel{(\ref{eq:balanced})}{\ge} &
1 - \O \left( \frac{\gamma_q(n)}{\sqrt{\N(d)}} \right) \\
\label{eq:balanced-k2}
& \stackrel{\mathrm{Lemma~\ref{lem:Ndelta}}}{=} &
1 - \O \left( \lambda_q(n) / \log_q n \right) .
\end{eqnarray}
Therefore,
\ifIEEE
    \begin{eqnarray*}
    \frac{|\calA_{k+d}|}{|\calA_k|}
    & \stackrel{(\ref{eq:balanced-k1})}{\ge} &
    \frac{\sum_{\bldtheta \in \calT}
       \left| \calA(\varphi(\bldtheta)) \right|}%
       {\sum_{\bldtheta \in \calT} \left| \calA(\bldtheta) \right|} \\
    &&
    \quad {}
    \cdot
    \left( 1 - \O \left( n^{17 - 8.5 \log_q n} \right) \right) \\
    & \stackrel{(\ref{eq:balanced-k2})}{\ge} &
    \left(
    1 - \O \left( \lambda_q(n) / \log_q n \right)
    \right) \\
    &&
    \quad {}
    \cdot
    \left( 1 - \O \left( n^{17 - 8.5 \log_q n} \right) \right) .
    \end{eqnarray*}
\else
\begin{eqnarray*}
\frac{|\calA_{k+d}|}{|\calA_k|}
& \stackrel{(\ref{eq:balanced-k1})}{\ge} &
\frac{\sum_{\bldtheta \in \calT} \left| \calA(\varphi(\bldtheta)) \right|}%
   {\sum_{\bldtheta \in \calT} \left| \calA(\bldtheta) \right|}
\cdot
\left( 1 - \O \left( n^{17 - 8.5 \log_q n} \right) \right) \\
& \stackrel{(\ref{eq:balanced-k2})}{\ge} &
\left(
1 - \O \left( \lambda_q(n) / \log_q n \right) \right)
\cdot
\left( 1 - \O \left( n^{17 - 8.5 \log_q n} \right) \right) .
\end{eqnarray*}
\fi
The result now follows by observing that
$n^{17 - 8.5 \log_q n} = \O \left( \lambda_q(n) / \log_q n \right)$.
\end{proof}

\begin{proof}[Proof of \Proposition~\ref{prop:flat}]
Let $k_0 = n - \deg \bldtheta_0$.
Then $\calA(\bldtheta_0) \subseteq \calA_{k_0}$ and, therefore,
$\calA_{k_0}$ is rich.
Recalling from~(\ref{eq:d_t=log_q(n)})
that $\delta \le \log_q (2n) + 1$ and that
\begin{eqnarray*}
\left| k - k_0 \right|
& \le &
\Bigl| k - \frac{\degb}{2} \Bigr|
+ \Bigl| k_0 - \frac{\degb}{2} \Bigr| \\
& \le &
\O \left( \log_q^2 n \right) + \frac{\rho_n}{2}
= \O \left( \log_q^2 n \right) ,
\end{eqnarray*}
we can write $k - k_0 = \pm (\ell \cdot \delta + c)$,
where $\ell$ and $c$ are nonnegative integers
and $\ell, c = \O(\log_q n)$.
For $j \in [1:\ell{+}2c]$, let
\[
k_j =
\left\{
\begin{array}{ll}
k_{j-1} \pm \delta
& \textrm{for $j \in [1:\ell{+}c]$} \\
k_{j-1} \mp (\delta - 1)
& \textrm{otherwise} ,
\end{array}
\right.
\]
where the sign in the first case is taken to match that of $k - k_0$
and is negated in the second case.
By $\ell + 2c$ repetitions of Lemma~\ref{lem:balanced-k}
we get inductively that
\begin{equation}
\label{eq:iterationk}
\frac{|\calA_{k_j}|}{|\calA_{k_0}|}
\ge
1 - \O \left( \lambda_q(n) \cdot j / \log_q n \right)
\end{equation}
and that $\calA_{k_j}$ is rich;
here we assume that~$n$ is above an absolute threshold
so that $\lambda_q(n)$ is sufficiently small
to guarantee that the right-hand side of~(\ref{eq:iterationk})
remains above, say, $0.8$. Now,
\[
k_{\ell+2c} = k_0 \pm (\ell+c) \delta \mp c(\delta - 1)
= k_0 \pm (\ell \delta + c) = k
\]
and, so,
\[
\frac{|\calA_k|}{|\calA_{k_0}|}
\ge 1 - \O \left( \lambda_q(n) \right) .
\]
By similar arguments we get that the last inequality holds
also when $(k,k_0)$ therein is replaced by $(k',k)$
(the constant $0.8$ makes $\calA_k$ sufficiently rich
to guarantee that all the traversed sets $\calA_{k_j}$ from
$\calA_k$ to $\calA_{k'}$ are rich).
\end{proof}

\begin{remark}
\label{rem:flat}
By a minor modification in the last proof, one can show that when
$\degb/\log_q^2 n$ is both $\Omega(1)$
and $o \left( 1 / \lambda_q(n) \right)$,
\Proposition~\ref{prop:flat} still holds
if the right-hand side of~(\ref{eq:flat}) is replaced by
$1 - \O \left( \lambda_q(n) \cdot \degb/\log_q^2 n \right)$.
\end{remark}
\fi

\section{Average-case analysis}
\label{sec:average-case}

\ifPAGELIMIT
    We start with two lemmas, the proofs of which we omit.
\else
We start with three lemmas.
\fi

\begin{lemma}
\label{lem:size-S_m}
For $m \in \Integers^+$ define the set
\[
\calS_m = \left\{ (a,b,c,d) \in \calP_m^4 \,:\,
\gcd(b,c) = 1 ,\ a b c d \in \calM_m \right\}.
\]
Then
\begin{equation}
\label{eq:size-S_m}
\left| \calS_m \right|
=
q^m \cdot
\left( \frac{q-1}{q}  \binom{m+1}{3} + (m+1)^2 \right)
.
\end{equation}
\end{lemma}

\ifPAGELIMIT
\else
\begin{proof}
Denote by $\calH_m$ the set
\[
\calH_m = \left\{(j,k,\ell) \in [0:m]^3 \,:\,
j+k+\ell\le m \right\} .
\]
For $(j,k,\ell) \in \calH_m$, let
\[
\calS_m(j,k,\ell) =
\bigl\{(a{,}b{,}c{,}d) \in \calS_m \,:\,
(a{,}b{,}c) \in
\calM_j \times \calM_k \times \calM_\ell
\bigr\} .
\]
By~\cite[\Theorem~3]{BenjaminBennett07}
it follows that when $k,\ell > 0$,
a fraction $(q-1)/q$ of the polynomial pairs in
$\calM_k \times \calM_\ell$ are relatively prime.
Hence,
\[
\left| \calS_m(j,k,\ell) \right| =
\left\{
\begin{array}{ccl}
(q-1)q^{m-1} && \textrm{if $k,\ell>0$} \\
q^m          && \textrm{if $k=0$ or $\ell=0$}
\end{array}
\right.
.
\]
Thus,
\begin{eqnarray*}
|\calS_m|
&=&
\sum_{(j,k,\ell) \in \calH_m} |\calS_m(j,k,\ell)| \\
& = &
\sum_{(j,k,\ell) \in \calH_m \,:\, k, \ell > 0}
|\calS_m(j,k,\ell)| \\
&&
\quad {} +
\sum_{(j,k,0) \in \calH_m \,:\, k>0}
|\calS_m(j,k,0)|
\\
&&
\quad {} +
\sum_{(j,0,\ell)\in\calH_m \,:\, \ell>0}
|\calS_m(j,0,\ell)|
\\
&&
\quad {} +
\sum_{j=0}^m|\calS_m(j,0,0)|
\\
& = &
q^m \cdot
\left(
\frac{q-1}{q}  \binom{m+1}{3}
+ 2 \binom{m+1}{2} + m+1
\right) \\
& = &
q^m \cdot
\left( \frac{q-1}{q}  \binom{m+1}{3} + (m+1)^2 \right)
.
\end{eqnarray*}
\end{proof}
\fi

\begin{lemma}
\label{lem:size-Sstar}
For $n \in \Integers^+$ define the set
\[
\calS_n^* = \left\{ (a{,}b{,}c{,}d) \in \calP_n^4 \,:\,
\gcd(b,c) = 1 ,\
a b, c d, a c, b d \in \calP_n
\right\} .
\]
Then
\begin{eqnarray*}
\left| \calS_n^* \right|
& = &
\frac{(n+1) q^{2n+1} (q+1)}{(q-1)^2}
- \frac{(q^{n+1} - 1)(3q^{n+1} - 1)}{(q-1)^3} \\
& = &
(n+1) q^{2n} \left( 1 + \O(1/q) \right) .
\end{eqnarray*}
\end{lemma}

\ifPAGELIMIT
\else
\begin{proof}
For any integer $t \ge 0$ define
\[
\phi(t) = \left| \left\{(b,c)\in \calP_t^2
\,:\, \gcd(b,c)=1 \right\} \right|.
\]
By~\cite[\Theorem~3]{BenjaminBennett07} it follows that
\begin{equation}
\label{eq:phi-value}
\phi(t)=\frac{q^{2t+1}-1}{q-1} .
\end{equation}
Now, for any fixed polynomials $a \in \calM_k$ and $d \in \calM_\ell$
(where $k, \ell \in [0:n]$), the quadruple $(a,b,c,d)$
is in $\calS_n^*$ \ifandonlyif\ $\gcd(b,c) = 1$ and
\[
\deg b, \deg c \le t = \min(n-k,n-\ell) .
\]
Hence,
\begin{eqnarray*}
|\calS_n^*|
& = &
\sum_{k=0}^n \sum_{\ell=0}^n q^k \cdot q^{\ell}
\cdot \phi(\min(n-k,n-\ell)) \nonumber\\
& = &
2\left( \sum_{k=0}^n \sum_{\ell=0}^k q^{k+\ell}
\phi(n-k) \right) - \sum_{k=0}^n q^{2k} \phi(n-k) \nonumber\\
\ifIEEE
    & \stackrel{(\ref{eq:phi-value})}{=} &
    2\left( \sum_{k=0}^n \sum_{\ell=0}^k q^{k+\ell}\cdot
    \frac{q^{2(n-k)+1}-1}{q-1}\right) \\
    &&
    \quad\quad {}
    -\sum_{k=0}^n q^{2k} \cdot \frac{q^{2(n-k)+1}-1}{q-1} ,
\else
& \stackrel{(\ref{eq:phi-value})}{=} &
2\left( \sum_{k=0}^n \sum_{\ell=0}^k q^{k+\ell}\cdot
\frac{q^{2(n-k)+1}-1}{q-1}\right)
-\sum_{k=0}^n q^{2k} \cdot \frac{q^{2(n-k)+1}-1}{q-1} ,
\fi
\end{eqnarray*}
where the second step follows by symmetry.
By simple algebra and summing the various geometric series,
we get the desired result.
\end{proof}

\begin{lemma}
\label{lem:size-Xstar}
For $n \in \Integers^+$ define the set
\begin{eqnarray*}
\calX^*_n  & = &
\Bigl\{ (f_j)_{j=1}^8 \in \calP_n^8 \,: \\
&& \quad\quad
f_1 {\cdot} f_2 {\cdot} f_3 {\cdot} f_4, \;
f_5 {\cdot} f_6 {\cdot} f_7 {\cdot} f_8, \\
&& \quad\quad\quad\quad
f_1 {\cdot} f_2 {\cdot} f_5 {\cdot} f_6, \;
f_3 {\cdot} f_4 {\cdot} f_7 {\cdot} f_8, \\
&& \quad\quad\quad\quad\quad\quad
f_1 {\cdot} f_3 {\cdot} f_5 {\cdot} f_7, \;
f_2 {\cdot} f_4 {\cdot} f_6 {\cdot} f_8 \in \calP_n , \\
&& \quad\quad
\gcd(f_3 {\cdot} f_4,f_5 {\cdot} f_6)
= \gcd(f_2 {\cdot} f_4,f_5 {\cdot} f_7) \\
&& \quad\quad\quad\quad
= \gcd(f_2,f_3) = \gcd(f_6,f_7) = 1
\Bigr\} .
\end{eqnarray*}
Then
\[
|\calX^*_n| = \O \left( n^4 \cdot q^{2n} \right) .
\]
\end{lemma}

\begin{proof}
For $m \in [0:2n]$, let $\calH_{m,n}$
be the set of all integer triples $\bldh = (h_1 \; h_2 \; h_3)$
such that
\[
h_i \in [0:n] \quad \textrm{and} \quad m-h_i \in [0:n] ,
\quad i = 1, 2, 3 .
\]
It is easy to see that
\begin{equation}
\label{eq:V}
|\calH_{m,n}| = \left( \min\{ m, 2n{-}m \} + 1 \right)^3 .
\end{equation}
For each $\bldh \in \calH_{m,n}$,
define the set $\calX_m(\bldh)$ by
\begin{eqnarray*}
\calX_m(\bldh)  & = &
\Bigl\{ (f_j)_{j=1}^8 \in \calP_m^8 \,: \\
&& \quad\quad
f_1 f_2 f_3 f_4 \in \calM_{h_1} ,   \;\;
f_5 f_6 f_7 f_8 \in \calM_{m-h_1} , \\
&& \quad\quad
f_1 f_2 f_5 f_6 \in \calM_{h_2} ,   \;\;
f_1 f_3 f_5 f_7 \in \calM_{h_3}
\Bigr\}
\end{eqnarray*}
(note that the elements of $\calX_m(\bldh)$
satisfy $\prod_{j=1}^8 f_j \in \calM_m$ and, so, we also have
$f_3 f_4 f_7 f_8 \in \calM_{m-h_2} \; (\subseteq \calP_n)$
and $f_2 f_4 f_6 f_8 \in \calM_{m-h_3} \; (\subseteq \calP_n)$).
It can be readily verified that
\begin{equation}
\label{eq:Xstar}
\calX_n^*
\subseteq
\bigcup_{m=0}^{2n} \bigcup_{\bldh \in \calH_{m,n}} \calX_m(\bldh) .
\end{equation}
Denoting $k_j = \deg f_j$,
the degree-lists $\bldk = (k_j)_{j=1}^8$ of the elements of
$\calX_m(\bldh)$ range over the solutions in
$[0:n]^8$ of the following set of linear equations:
\begin{equation}
\label{eq:k}
\left(
\begin{array}{cccccccc}
1 & 1 & 1 & 1 & 0 & 0 & 0 & 0 \\
0 & 0 & 0 & 0 & 1 & 1 & 1 & 1 \\
1 & 1 & 0 & 0 & 1 & 1 & 0 & 0 \\
1 & 0 & 1 & 0 & 1 & 0 & 1 & 0
\end{array}
\right)
\bldk =
\left(
\begin{array}{c}
h_1 \\
m - h_1 \\
h_2 \\
h_3
\end{array}
\right) .
\end{equation}
Since the matrix has full rank, the number of such solutions
is bounded from above by $(n+1)^4$. Hence,
\begin{equation}
\label{eq:Xmh}
|\calX_m(\bldh)| \le (n+1)^4 \cdot q^m .
\end{equation}
Summarizing,
\begin{eqnarray*}
|\calX_n^*|
& \stackrel{(\ref{eq:Xstar})}{\le} &
\sum_{m=0}^{2n}
\sum_{\bldh \in \calH_{m,n}} |\calX_m(\bldh)| \\
& \stackrel{(\ref{eq:Xmh})}{\le} &
(n+1)^4 \sum_{t=0}^{2n} |\calH_{2n-t,n}| \cdot q^{2n-t} \\
& \stackrel{(\ref{eq:V})}{\le} &
(n+1)^4 \cdot q^{2n}
\cdot \sum_{t=0}^{2n} (t{+}1)^3 q^{-t} \\
& = &
\O(n^4 \cdot q^{2n}) .
\end{eqnarray*}
\end{proof}
\fi

\begin{proof}%
        [Proof of \Theorem~\ref{thm:expectation-variance-unconstrained}]
\ifPAGELIMIT
    Starting with $\Expected \left\{ \random_m \right\}$,
    for $s(x) \in \calM_m$,
\else
We start with the expectation of $\random_m$.
For each $s(x) \in \calM_m$,
\fi
let
\[
\calJ(s) =
\left\{
(u(x),v(x)) \in \calP_m^2 \,:\, s(x) = u(x) v(x)
\right\} .
\]
We have:
\ifPAGELIMIT
    \begin{eqnarray*}
    q^m \cdot \Expected \left\{ \random_m \right\}
    \!
    & \!=\! &
    \textstyle
    \sum_{s \in \calM_m} |\calJ(s)|
    = \bigl| \bigcup_{k=0}^m
    \left( \calM_k \times \calM_{m-k} \right) \bigr| \\
    & \!=\! &
    \textstyle
    \sum_{k=0}^m
    \left| \calM_k \right| \cdot \left| \calM_{m-k} \right|
    = (m+1) \cdot q^m .
   \end{eqnarray*}
\else
\begin{eqnarray*}
q^m \cdot \Expected \left\{ \random_m \right\}
=
\sum_{s \in \calM_m} |\calJ(s)|
& \!\!=\!\! &
\Bigl| \bigcup_{k=0}^m
\left( \calM_k \times \calM_{m-k} \right) \Bigr| \\
& \!\!=\!\! &
\sum_{k=0}^m
\left| \calM_k \right| \cdot \left| \calM_{m-k} \right| \\
& \!\!=\!\! &
(m+1) \cdot q^m .
\end{eqnarray*}
\fi

Turning to
\ifPAGELIMIT
    $\Variance \{ \random_m \}$,
\else
the variance of $\random_m$,
\fi
we define the set
\[
\calQ_m =
\left\{
(u,v,\hat{u},\hat{v}) \in \calP_m^4 \,:\,
u(x) v(x) = \hat{u}(x) \hat{v}(x) \in \calM_m
\right\}.
\]
It is easy to see that
\begin{equation}
\label{eq:variance-unconstrained}
\ifPAGELIMIT
    \textstyle
\fi
\left|\calM_m\right|\cdot \Expected \left\{ \random_m^2 \right\}
= \sum_{s \in \calM_m} |\calJ(s)|^2
= \left| \calQ_m \right|.
\end{equation}
Let $\calS_m$ be as in Lemma~\ref{lem:size-S_m},
and consider the mapping from $\calS_m$ to $\calQ_m$ that sends
each quadruple $(a,b,c,d) \in \calS_m$ to
a quadruple $(u,v,\hat{u},\hat{v}) \in \calQ_m$ by
\begin{equation}
\label{eq:StoQ}
u = a b,\; v = c d,\; \hat{u} = a c,\; \hat{v} = b d .
\end{equation}
Under this mapping,
each quadruple $(u,v,\hat{u},\hat{v}) \in\calQ_m$ is an image
of a (unique) quadruple  $(a,b,c,d) \in \calS_m$ given by
\[
a = \gcd(u,\hat{u}),\;
d = \gcd(v,\hat{v}),\;
\ifPAGELIMIT
    b = u/a,\;
    c = v/d,
\else
b = \frac{u}{a}=\frac{\hat{v}}{d},\;
c = \frac{v}{d}=\frac{\hat{u}}{a} .
\fi
\]
\ifPAGELIMIT
    i.e., (\ref{eq:StoQ}) defines a bijection
    $\calS_m \rightarrow \calQ_m$ and, so,
    $\left| \calQ_m \right| = \left| \calS_m \right|$.
\else
Hence, (\ref{eq:StoQ}) defines a bijection from~$\calS_m$ to~$\calQ_m$
and, so,
\begin{equation}
\label{eq:QS}
\left| \calQ_m \right| = \left| \calS_m \right| .
\end{equation}
\fi
Combining with (\ref{eq:size-S_m}) and~(\ref{eq:variance-unconstrained})
finally yields
\[
\Variance \{ \random_m \}
= \Expected \left\{ \random_m^2 \right\}
- \left(\Expected \left\{ \random_m \right\} \right)^2
= \frac{q-1}{q}\cdot \binom{m+1}{3}.
\]
\end{proof}

\ifPAGELIMIT
    The proof of \Proposition~\ref{prop:chernoff-tail} is omitted.

    The proof of \Theorem~\ref{thm:expectation-variance-constrained}
    is similar to the computation of the variance in the proof
    of \Theorem~\ref{thm:expectation-variance-unconstrained}:
    we use Lemma~\ref{lem:size-Sstar} and a refinement of it
    to compute (\respectively, bound)
    the expectation (\respectively, variance) of $\random_{n,n}$.
\else
\begin{proof}[Proof of
\Theorem~\ref{thm:expectation-variance-constrained}]
We start with the expectation of $\random_{n,n}$.
For each $(u,v) \in \calP_n^2$, define
\[
\calL(u,v) =
\left\{ (\hat{u},\hat{v}) \in \calP_n^2 \,:\,
u \cdot v = \hat{u} \cdot \hat{v} \right\}
\]
and let
\begin{eqnarray*}
\calQ_n^*
& = &
\left\{
(u,v,\hat{u},\hat{v}) \in \calP_n^4 \,:\,
u(x) v(x) = \hat{u}(x) \hat{v}(x) \right\} \\
& = &
\bigcup_{(u,v) \in \calP_n^2}
\left\{
(u,v,\hat{u},\hat{v}) \in \calP_n^4 \,:\,
(\hat{u},\hat{v}) \in \calL(u,v)
\right\} .
\end{eqnarray*}
Then,
\[
\left| \calP_n^2 \right| \cdot \Expected \left\{ \random_{n,n} \right\}
= \sum_{(u,v) \in \calP_n^2} |\calL(u,v)|
= \left| \calQ_n^* \right| .
\]
We now apply essentially the same arguments
that lead to the equality~(\ref{eq:QS}).
We re-define the mapping~(\ref{eq:StoQ}) to be from $\calS_n^*$
to $\calQ_n^*$
(where $\calS_n^*$ was defined in Lemma~\ref{lem:size-Sstar});
by~(\ref{eq:StoQ}), this mapping is a bijection and, so,
$\left| \calQ_n^* \right| = \left| \calS_n^* \right|$

In summary, we have shown that
$|\calP_n|^2 \cdot \Expected \left\{ \random_{n,n} \right\}
= \left| \calS_n^* \right|$ which,
with $|\calP_n| = (q^{n+1} - 1)/(q-1)$
and Lemma~\ref{lem:size-Sstar}, yields:
\begin{eqnarray*}
\Expected \left\{ \random_{n,n} \right\}
& = &
\frac{|\calS_n^*|}{|\calP_n|^2} \\
& = &
(n+1) \cdot \frac{q^{2n+1}(q+1)}{(q^{n+1} - 1)^2}
- \frac{3q^{n+1}-1}{(q^{n+1}-1)(q-1)} \\
& = &
(n+1)(1+\O(1/q)) .
\end{eqnarray*}

We now turn to bounding from above the variance of $\random_{n,n}$.
It is straightforward to see that
\[
|\calL(u,v)|^2
= \left|\left\{(u_1,v_1,u_2,v_2) \in \calP_n^4 \,:\,
u_1 v_1 = u_2 v_2 = u v \right\} \right|.
\]
Restricting the bijection from $\calS_n^*$ to $\calQ_n^*$,
which is defined by~(\ref{eq:StoQ}) to a domain
where the products $a b$ and $c d$ are fixed to be~$u$ and~$v$,
respectively, the range becomes the set of
all quadruples $(u,v,\hat{u},\hat{v})$ such that
$(\hat{u},\hat{v}) \in \calL(u,v)$. Hence,
\begin{eqnarray*}
|\calL(u,v)|^2 & = &
\bigl|\bigl\{(a_1,b_1,c_1,d_1,a_2,b_2,c_2,d_2)\in\calP_n^8 \,: \\
&&
\quad
\ifIEEE
\else
\quad
\fi
(a_1,b_1,c_1,d_1), (a_2,b_2,c_2,d_2) \in \calS_n^* , \\
&&
\quad\quad
\ifIEEE
\else
\quad\quad
\fi
u = a_1 b_1 = a_2 b_2 ,\
v = c_1 d_1 = c_2 d_2
\bigr\}\bigr|.
\end{eqnarray*}
Defining
\begin{eqnarray*}
\calE_n^* & = &
\bigl\{(a_1,b_1,c_1,d_1,a_2,b_2,c_2,d_2) \in \calP_n^8 \,: \\
&&
\quad
\ifIEEE
\else
\quad
\fi
(a_1,b_1,c_1,d_1), (a_2,b_2,c_2,d_2) \in \calS_n^* , \\
&&
\quad\quad
\ifIEEE
\else
\quad\quad
\fi
a_1 b_1 = a_2 b_2,\; c_1 d_1 = c_2 d_2
\bigr\},
\end{eqnarray*}
we therefore have
\begin{equation}
\label{eq:|Q|-and-2nd-moment}
\left| \calP_n^2 \right| \cdot \Expected\left\{\random_{n,n}^2\right\}
= \sum_{(u,v) \in \calP_n^2} |\calL(u,v)|^2
=|\calE_n^*|.
\end{equation}

We next give an upper bound on $|\calE_n^*|$
using Lemma~\ref{lem:size-Xstar}.
Similarly to the arguments that lead to (\ref{eq:QS}),
we observe that
the following mapping from~$\calX^*_n$ to~$\calE^*_n$
is a bijection:
\[
\arraycolsep0.2ex
\renewcommand{\arraystretch}{1.3}
\begin{array}{rclrclrclrcl}
a_1 & = & f_1 f_2, \;\; &
b_1 & = & f_3 f_4, \;\; &
c_1 & = & f_5 f_6, \;\; &
d_1 & = & f_7 f_8,      \\
a_2 & = & f_1 f_3,      &
b_2 & = & f_2 f_4,      &
c_2 & = & f_5 f_7,      &
d_2 & = & f_6 f_8 .
\end{array}
\]
Hence,
\[
|\calE_n^*| = \O \left( n^4 \cdot q^{2n} \right) ,
\]
which, with~(\ref{eq:|Q|-and-2nd-moment})
and $|\calP_n| = (q^{n+1} - 1)/(q-1)$, yields
$\Expected \left \{ \random_{n,n}^2 \right\} = \O(n^4)$.
Since $\Expected \left\{ \random_{n,n} \right \} = \O(n)$,
we conclude that
\[
\Variance \left\{ \random_{n,n} \right\}
= \Expected\left\{\random_{n,n}^2\right\}-
\left( \Expected \left \{ \random_{n,n} \right\} \right)^2 = \O(n^4).
\]
\end{proof}

\begin{remark}
\label{rem:expectation-variance-constrained}
The $\O(n^4)$ expression for
$\Variance \left\{ \random_{n,n} \right\}$
in \Theorem~\ref{thm:expectation-variance-constrained}
can be tightened to $\Theta(n^4)$, at least for $q \ge 9$.
To see this, we note that a containment (rather than equality)
holds in~(\ref{eq:Xstar}) since we disregard
the constraints in the definition of $\calX_n^*$
that certain pairs of polynomials $(f_i,f_j)$ should be
relatively prime. Specifically, in that definition,
we require that $\gcd(f_i,f_j) = 1$ for the following
nine pairs $(i,j)$:
\[
(2,3), (2,5), (2,7), (3,5), (3,6), (4,5), (4,6), (4,7), (6,7) .
\]
In this list, we can find three pairs that are disjoint, say,
$(2,3)$, $(4,5)$, and $(6,7)$.
By~\cite[\Theorem~3]{BenjaminBennett07} it then follows that
for every $\bldh \in \calH_{m,n}$ and $q \ge 9$,
\[
|\calX_n^*|
\ge
\left( \Bigl( \frac{q-1}{q} \Bigr)^3 - \frac{6}{q} \right)
\cdot |\calX_m(\bldh)| > 0.03 \cdot |\calX_m(\bldh)| .
\]
This holds in particular for
$\bldh = n \cdot (1 \; 1 \; 1)$, which belongs to $\calH_{2n,n}$.
For this~$\bldh$ we have $|\calX_m(\bldh)| = \Theta(n^4 \cdot q^{2n})$,
since we can exhibit $\Theta(n^4)$ solutions~$\bldk$
for~(\ref{eq:k}):
\[
\bldk = (k_j)_{j=1}^8 + \Lambda^\top \blda ,
\]
where
\begin{eqnarray*}
k_1 = k_2 = k_3 = k_5 & = & \floor{(n{+}2)/4} , \\
k_4 = k_6 = k_7       & = & n - 3 \floor{(n{+}2)/4} , \\
k_8                   & = & 5 \floor{(n{+}2)/4} - n ,
\end{eqnarray*}
$\blda = (a_i)_{i=1}^4$ is any column vector in $\Integers^4$
that satisfies $\sum_{i=1}^4 |a_i| \le n/4 - 2$, and
\[
\Lambda
=
\left(
\begin{array}{cccccccc}
+ & - & + & - & - & + & - & + \\
+ & + & - & - & - & - & + & + \\
+ & - & - & + & + & - & - & + \\
+ & - & - & + & - & + & + & -
\end{array}
\right) ,
\]
with ``$+$'' and ``$-$'' standing for~$1$ and~$-1$, respectively
(the rows of~$\Lambda$ span the right kernel of
the matrix in~(\ref{eq:k})).\qed
\end{remark}
\fi

\ifPAGELIMIT
\else
\ifIEEE
   \appendices
\else
   \ifJCTA
   \else
   \section*{$\,$\hfill Appendices\hfill$\,$}
    \fi
   \appendix
\fi

\section{Proof of \Proposition~\protect\ref{prop:chernoff-tail}}
\label{sec:chernoff-tail}

We will make use of the following known bound.

\begin{theorem}[Chernoff bound~{\cite[p.~127]{Gallager}}]
\label{thm:chernoff}
Given a random variable~$X$,
for every real~$w$ and $\alpha > 1$:
\[
\Prob \left\{ X \ge w \right\}
\le \alpha^{-w} \cdot \Expected \left\{ \alpha^X \right\} .
\]
\end{theorem}

We assume a uniform distribution on $\calM_m$ and define
a random variable $\X_m : \calM_m \rightarrow \Integers$
which maps each $s(x) \in \calM_m$ to the number of
irreducible factors of $s(x)$ over $\F$ (counting multiplicities).
Our proof of \Proposition~\ref{prop:chernoff-tail}
will be based on the following inequality, which holds for
every real~$\beta$:
\begin{eqnarray}
\Prob \left\{ \random_m \ge m^\beta \right\}
& \le &
\Prob \left\{ 2^{\X_m} \ge m^\beta \right\} \nonumber \\
\label{eq:twoprobs}
& = &
\Prob \left\{ \X_m \ge \beta \log_2 m \right\} .
\end{eqnarray}

Let $P(z,u)$ denote the bivariate generating function
of the number of polynomials in $\calM_m$ that have~$k$
monic irreducible factors (counting multiplicity). Then:
\[
P(z,u)
= \sum_{m=0}^\infty \sum_{k=0}^\infty
q^m \cdot \Prob \left\{ \X_m = k \right\} z^m u^k .
\]
On the other hand, we also have~\cite[Eq.~(10)]{FGP}:
\[
P(z,u) = \prod_{d=1}^\infty (1 - u z^d)^{-\I(d)} ,
\]
namely,
\begin{eqnarray*}
\ln P(z,u)
& = & -\sum_{d=1}^\infty \I(d) \ln (1 - u z^d) \\
& = & \sum_{d=1}^\infty \I(d) \sum_{k=1}^\infty \frac{(u z^d)^k}{k} \\
& = & \sum_{d=1}^\infty d \, \I(d)
\sum_{k=1}^\infty \frac{u^k z^{d k}}{d k} \\
& = & \sum_{m=1}^\infty \frac{(q z)^m}{m} \, G_m(u) ,
\end{eqnarray*}
where
\begin{equation}
\label{eq:Gn}
G_m(u) = \frac{1}{q^m} \sum_{d \in \Integers^+ \,:\, d \,|\, m}
d \, \I(d) \cdot u^{m/d} .
\end{equation}
Hence, for every $\alpha > 1$:
\begin{equation}
\label{eq:Pzalpha}
P(z/q,\alpha)
= \sum_{m=0}^\infty
\Expected \left\{ \alpha^{\X_m} \right\} z^m =
\exp \left\{ \sum_{m=1}^\infty
\frac{z^m}{m} \, G_m(\alpha) \right\} .
\end{equation}

We will limit ourselves to~$\alpha$ in the interval $(1,q)$.
Denote
\[
\varepsilon_m = \frac{G_m(\alpha) - \alpha}{m} .
\]
By~(\ref{eq:Gn}) we have,
for every $\alpha \in (1,q)$:
\begin{eqnarray*}
\ifIEEE
    \sum_{m=1}^\infty |\varepsilon_m|
    & \le &
    \sum_{m=1}^\infty
    \frac{1}{m}
    \biggl(
    q \cdot \frac{\alpha^m}{q^m} \\
    && \quad {}
    + \sum_{1 < d < m \,:\, d \,|\, m}
    \frac{d \, \I(d) \cdot \alpha^{m/d}}{q^m} \\
    && \quad {}
    + \alpha \cdot \frac{\left| m \, \I(m) - q^m \right|}{q^m}
    \biggr)
\else
\sum_{m=1}^\infty |\varepsilon_m|
\!
& \le &
\!
\sum_{m=1}^\infty
\frac{1}{m}
\biggl(
q \cdot \frac{\alpha^m}{q^m} 
+ \!\! \sum_{1 < d < m \,:\, d \,|\, m} \!\!\!
\frac{d \, \I(d) \cdot \alpha^{m/d}}{q^m}
+ \alpha \cdot \frac{\left| m \, \I(m) - q^m \right|}{q^m}
\biggr)
\fi
\\
\ifIEEE
    & \stackrel{(\ref{eq:bounds-on-I(d)})}{<} &
    q \sum_{m=1}^\infty \frac{1}{m} \cdot \frac{\alpha^m}{q^m} \\
    && \quad{}
    + \alpha^2 \sum_{m=4}^\infty q^{-m/2} \\
    && \quad{}
    + \alpha \sum_{m=2}^\infty q^{-m/2}
\else
& \stackrel{(\ref{eq:bounds-on-I(d)})}{<} &
q \sum_{m=1}^\infty \frac{1}{m} \cdot \frac{\alpha^m}{q^m}
+ \alpha^2 \sum_{m=4}^\infty q^{-m/2}
+ \alpha \sum_{m=2}^\infty q^{-m/2}
\fi
\\
& = &
- q \ln \left( 1 - \frac{\alpha}{q} \right)
+ \frac{(\alpha/q)^2 + (\alpha/q)}{1 - \sqrt{1/q}} \\
& < &
- q \ln \left( 1 - \frac{\alpha}{q} \right)
+ \frac{2}{1 - \sqrt{1/q}} .
\end{eqnarray*}
Hence,
\[
\sigma(\alpha)
=
\exp \left\{ \sum_{m=1}^\infty |\varepsilon_m| \right\}
= \O \left( \frac{1}{(1 - (\alpha/q))^q} \right)
< \infty
\]
(where the constant in the $\O(\cdot)$ term is absolute).

\begin{lemma}
\label{lem:moment}
For every $\alpha \in (1,q)$ and $m \in \Integers^+$,
\[
\frac{\Expected \left\{ \alpha^{\X_m} \right\}}{m^{\alpha-1}}
\le \frac{\sigma(\alpha)}{\Gamma(\alpha)}
\cdot \left( 1 + o_m(1) \right) ,
\]
where $\Gamma(\cdot)$ denotes the Gamma function
and $o_m(1)$ stands for an expression that goes to~$0$
as $m \rightarrow \infty$ (uniformly over $\alpha \in (1,q)$).
\end{lemma}

\begin{proof}
   From~(\ref{eq:Pzalpha}) we get:
\begin{eqnarray*}
\sum_{m=1}^\infty
\Expected \left\{ \alpha^{\X_m} \right\} z^m
& \!\!\!=\!\!\! &
\exp \left\{ \alpha \sum_{m=1}^\infty
\frac{z^m}{m} \right\} \cdot
\exp \left\{ \sum_{m=1}^\infty \varepsilon_m z^m \right\} \\
& \!\!\!=\!\!\! &
\frac{1}{(1-z)^\alpha} \cdot
\exp \left\{ \sum_{m=1}^\infty \varepsilon_m z^m \right\} .
\end{eqnarray*}
Write
\begin{eqnarray*}
\frac{1}{(1-z)^\alpha}
& = & \sum_{m=0}^\infty f_m z^m , \\
\exp \left\{ \sum_{m=1}^\infty \varepsilon_m z^m \right\}
& = & \sum_{m=0}^\infty g_m z^m , \\
\exp \left\{ \sum_{m=1}^\infty |\varepsilon_m| z^m \right\}
& = & \sum_{m=0}^\infty h_m z^m .
\end{eqnarray*}
Then
\[
f_m = \frac{\alpha(\alpha+1)(\alpha+2)\cdots(\alpha+m-1)}{m!} ,
\]
which is an increasing sequence in~$m$.
We recall that one the definitions of the Gamma function is
the following limit~\cite[p.~3]{AAR}:
\[
\Gamma(\alpha) = \lim_{m \rightarrow \infty} \frac{m^{\alpha-1}}{f_m}
\]
(where the convergence is uniform over $\alpha \in [1,q]$).
Since the series expansion of $\exp \{ \cdot \}$
contains only positive coefficients, it readily
follows that $g_m \le h_m$ for all~$m$. Hence,
\begin{eqnarray*}
\Expected \left\{ \alpha^{\X_m} \right\}
& = &
\sum_{i=0}^m f_{m-i} g_i
\le
\sum_{i=0}^m f_{m-i} h_i
\le
f_m \sum_{i=0}^\infty h_i \\
& = &
\sigma(\alpha) \cdot f_m .
\end{eqnarray*}
The result follows.
\end{proof}

Applying \Theorem~\ref{thm:chernoff}
to $X = \X_m$, $w = c \ln m$, and $\alpha \in (1,q)$
yields the upper bound
\begin{eqnarray}
\Prob \left\{ \X_m \ge c \ln m \right\}
& \le &
\alpha^{-c \ln m} \cdot \Expected \left\{ \alpha^{\X_m} \right\}
\nonumber \\
\label{eq:chernoff}
& = & \O \left( m^{-c \ln \alpha + \alpha - 1} \right) ,
\end{eqnarray}
where the last step follows from Lemma~\ref{lem:moment},
and the constant in the $\O(\cdot)$ term is absolute
if~$\alpha$ is a constant independent of~$q$.
For a given~$c < q$, the power of~$m$ in~(\ref{eq:chernoff})
attains its minimum over $\alpha \in [1,q)$
for $\alpha = \max \{ c, 1 \}$.
Therefore, that power is negative \ifandonlyif\ $c > 1$.
Combining with~(\ref{eq:twoprobs}),
we thus have the following upper bound
for every $c \in (1,q)$:
\[
\Prob \left\{ \random_m \ge m^{c \ln 2} \right\}
\le \O \left( m^{-c \ln c + c - 1} \right) ,
\]
where the constant in the $\O(\cdot)$ term is absolute
for constant~$c$. In particular,
taking $c = 1 + (\varepsilon/\ln 2)$
yields \Proposition~\ref{prop:chernoff-tail}.
\fi

\end{document}